\documentclass[aps,prd,twocolumn,showpacs,longbibliography]{revtex4-1}

\usepackage{amsmath,amssymb,amsfonts}
\usepackage{dsfont}
\usepackage{CJK}
\usepackage{CJKpunct}

\usepackage{graphicx}
\usepackage{xcolor}

\usepackage[CJKbookmarks, pdftex, bookmarksnumbered, bookmarksopen, colorlinks, citecolor=blue, linkcolor=blue]{hyperref}
\AtBeginDvi{}
\usepackage{verbatim} 

\newcommand{\Rl}[1]{\mathrm{Re}(#1)}                
\newcommand{\p}{\partial}                           
\newcommand{\dd}{\mathrm{d}}                        
\newcommand{\im}{\mathrm{i}}                        
\newcommand{\ex}{\mathrm{e}}                        
\newcommand{\tr}{\mathrm{tr}}                       
\newcommand{\ket}[1]{|{#1}\rangle}                  
\newcommand{\bra}[1]{\langle{#1}|}                  
\newcommand{\braket}[2]{\langle{#1}|{#2}\rangle}    

\newcommand{\D}{\mathcal{D}}                        

\CJKtilde   

\begin{document}
  \title{Euclidean three-point function in loop and perturbative gravity}
  \author{Carlo Rovelli}
  \email{rovelli@cpt.univ-mrs.fr}
   \affiliation{Centre de Physique Th\'eorique de Luminy\footnote{Unit\'e mixte de recherche du CNRS et des Universit\'es de Provence, de la M\'editerran\'ee et du Sud; affili\'e \`a la FRUMAN.},      Case 907, F-13288 Marseille, EU     }
   \author{Mingyi Zhang}
   \email{Mingyi.Zhang@cpt.univ-mrs.fr}
   \affiliation{Centre de Physique Th\'eorique de Luminy\footnote{Unit\'e mixte de recherche du CNRS et des Universit\'es de Provence, de la M\'editerran\'ee et du Sud; affili\'e \`a la FRUMAN.},      Case 907, F-13288 Marseille, EU     }
  \date{\today}

  \begin{abstract}
\noindent We compute the leading order of the three-point  function in loop quantum gravity, using the vertex expansion of the Euclidean version of the new spin foam dynamics, in the region of $\gamma<1$.  We find results consistent with Regge calculus in the limit $\gamma\rightarrow0$, $j\rightarrow\infty$.  We also compute the tree-level three-point function of perturbative quantum general relativity in position space, and discuss the possibility of directly comparing the two results.
   \end{abstract}
  \maketitle

  \section{Introduction}

The difficulty of extracting physical predictions from a background-independent theory is a well-known difficulty of quantum gravity.  A strategy to address the problem has been developing in recent years, based on two ideas.  The first is to define $n$-point functions over a background by storing the information about the background in the boundary state \cite{Modesto:2005sj}.  In covariant loop gravity \cite{Rovelli:QG,Rovelli:2011eq}, this technique yields a definite expression for the theory's $n$-point functions. The second  is to explore the expansion of this expression order by order in the number of interaction vertices \cite{Rovelli:2005yj}.  Although perhaps counter-intuitive, this expansion has proven effective in certain regimes; for details see  \cite{Bianchi:2006uf,Speziale:2008uw}.   In particular, the low-energy limit of the two-point  function (the ``graviton propagator") obtained in this way from the improved-Barrett-Crane spin foam dynamics \cite{Engle:2007uq,Livine:2007vk,Engle:2007qf,Pereira:2007nh,Freidel:2007py,Engle:2007wy}  (sometime denoted the EPRL/FK model)  correctly matches the graviton propagator of pure gravity in a transverse radial gauge (harmonic gauge) \cite{Alesci:2008ff,Bianchi:2009ri}. This result has been possible thanks to the introduction of the coherent intertwiner basis \cite{Livine:2006it} and the asymptotic analysis of vertex amplitude \cite{Barrett:2009gg,Barrett:2009mw}.

The obvious next step is to compute the three-point  function. In this paper we begin the three-point  function analysis.  We compute the three-point  function from the non-perturbative theory.  As in  \cite{Bianchi:2009ri}, we work in the Euclidean regime and with the Barbero-Immirzi parameter $0<\gamma<1$ where the amplitude defined in \cite{Freidel:2007py} and that defined in   \cite{Engle:2007wy} coincide.

Our main  result is the following.  We consider the limit, introduced in \cite{Bianchi:2009ri}, where the Barbero-Immirzi parameter is taken to zero $\gamma\rightarrow0$, and the spin of the boundary state is taken to infinity $j\rightarrow\infty$, keeping the size of the quantum geometry  $A\sim\gamma j$ finite and fixed. This limit corresponds to  neglecting Planck scale discreteness effects, at large finite distances.  In this limit, the three-point  function we obtain exactly matches the one obtained from Regge calculus \cite{Regge:1961px}.

This implies that the spin foam dynamics  is consistent with a discretization of general relativity, not just in the quadratic approximation, but also to the first order in the interaction terms.

The relation between the Regge and Loop three-point  function and the three-point  function of the weak field perturbation expansion of general relativity around flat space, on the other hand, remains elusive. We compute explicitly the perturbative three-point  function in position space in the transverse gauge (harmonic gauge), and we discuss the technical difficulty of comparing this with the  Regge/Loop one.

The paper is organized as follow: in Section \ref{sec:3SF}, we review the ingredients and the assumptions needed to to define the $n$-point functions in loop gravity and we compute the three-point function.  In Section \ref{sec:3p} we derive the three-point  function  from perturbative field theory, and discuss the relation between this and the Loop/Regge one.

  \section{Three-point function in loop gravity}\label{sec:3SF}

In this section we compute the three-point  function of the spin foam amplitude in loop quantum gravity at first order in the vertex expansion.  We follow closely the techniques developed for the two-point function in \cite{Bianchi:2006uf,Bianchi:2009ri} and the calculation of the three-point  function for the old Barrett-Crane model in \cite{Bianchi:RC2008}. For previous work in this direction, see also \cite{Alesci:2007tx,Alesci:2007tg,Alesci:2008ff}.

 \subsection{Boundary Formalism}

The well known difficulty of defining $n$-point functions in a general covariant quantum field theory can be illustrated by the following (naive) argument. If the action $S[g]$ and the measure are invariant under coordinate transformations, then
  \begin{equation}\label{eq:ZQFT}
    W(x_1,\cdots,x_N)\sim \int \D g \ g(x_1)\cdots g(x_N)\ \ex^{\im S[g]}
  \end{equation}
is formally independent from $x_n$ (as long as the $x_n$ are distinct), because a change in $x_n$ can be absorbed into a change of coordinates that leaves the integral invariant.  This difficulty is circumvented in the weak field approximation as follows.  If we want to study the theory around flat space, we have to impose boundary conditions on Eq.\eqref{eq:ZQFT} demanding that $g$ goes to flat space at infinity.  With this choice, the classical solution that dominates the path integral in the weak field limit is flat spacetime. In flat spacetime, we can choose preferred Cartesian coordinates $x$, and write the field insertions in terms of \emph{these} preferred coordinates.  Then Eq.\eqref{eq:ZQFT}  is well defined: the coordinates $x_n$ are not generally covariant coordinates, but rather Minkowski coordinates giving physical distances and physical time intervals in the background metric picked out by the boundary conditions of the field at infinity. This is the way $n$-point functions are defined for perturbative  general relativity. In the full non-perturbative theory, on the other hand, this strategy is not viable, because the integral Eq.\eqref{eq:ZQFT} has formally to be taken over arbitrary geometries, where the notion of preferred Cartesian coordinate loses meaning.

The idea for solving this difficulty was introduced in \cite{Modesto:2005sj} and is explained in detail in \cite{Bianchi:2006uf}. We give here a short account of this formalism, but we urge the reader to look at the original references for a detailed explanation of the approach. Let us begin by picking a surface $\Sigma$ in flat spacetime, bounding a compact region $\cal R$, and approximate Eq.\eqref{eq:ZQFT} by replacing $S[g]$  outside $\cal R$ with the linearized action.
Then split Eq.\eqref{eq:ZQFT} into three integrals: the integral on the field variables in $\cal R$, outside $\cal R$, and on $\Sigma$. Let $\gamma$ be the value of the field on $\Sigma$.  Let $W_\Sigma[\gamma]$ be the result of the internal integration, at fixed value $\gamma$ of the field on $\Sigma$
  \begin{eqnarray}\label{eq:W}
    W_\Sigma[\gamma] = \int_{g|_\Sigma=\gamma}\D g\ ~ \ex^{\im S[g]}\ .
 \end{eqnarray}
Let  $\Psi_\Sigma[\gamma]$ be the result of the outside integral. Then we can write
  \begin{eqnarray}\label{eq:QG}\nonumber
 W(x_1, ... x_N)&\sim&
\int\D\gamma\ ~W_\Sigma[\gamma]\ \gamma(x_1)...\gamma(x_N)
 \Psi_\Sigma[\gamma]\\ &\equiv& \langle  W_\Sigma  |  \gamma(x_1)...\gamma(x_N)|   \Psi_\Sigma \rangle
 \end{eqnarray}
Now observe first that because of the (assumed) diff-invariance of measure and action, $W_\Sigma[\gamma]$ is in fact independent from $\Sigma$. That is $W_\Sigma=W$. Second,  since the external integral is that of a free theory, $ \Psi_\Sigma[\gamma]$, will be the vacuum state of the free theory on the surface $\Sigma$. This can be shown to be a Gaussian semiclassical state peaked on the intrinsic and extrinsic geometry of $\Sigma$. Inserting the proper normalization we write
  \begin{equation}\label{eq:QG}
 W(x_1, ... x_N) = \langle \gamma(x_1)...\gamma(x_N) \rangle= \frac{\langle  W  |  \gamma(x_1)...\gamma(x_N)|   \Psi_\Sigma \rangle}{\langle  W  | \Psi_\Sigma \rangle}
 \end{equation}
where $W$ is the formal functional integral on a compact region, and $\Psi_\Sigma$ is a semiclassical state peaked on a certain intrinsic and extrinsic geometry.  This is the ``boundary formalism". For a strictly related approach, see also \cite{Oeckl:2005bv,Oeckl:2003vu}.  The quantities appearing in the formal expression Eq.\eqref{eq:QG} are well defined in loop quantum gravity and this expression can be taken as the starting point for computing $n$-point functions from the background independent theory.

\subsection{The theory}\label{sec:TH}

The definition of the non perturbative quantum gravity theory we use is given for instance in \cite{Rovelli:2011eq}. The Hilbert space of the theory is spanned by spin network states $|\Gamma,\psi\rangle$, where $\Gamma$ is a graph with $L$ links $l$ and $N$ nodes $n$ and $\psi$ is in ${\cal H}_\Gamma=L_2[SU(2)^L/SU(2)^N]$. A convenient basis in ${\cal H}_\Gamma$ is given by the coherent states $|j,\vec n\rangle$ which are the gauge invariant projections of  $SU(2)$ Bloch coherent states \cite{Livine:2007vk}. These are labeled by a spin $j_l$ per each link of the graph, and a unit-norm 3-vector $\vec n_{nl}$ for each couple node-link of the graph.  The dynamics of the theory is determined by the amplitude $W$ defined as a sum over two-complexes, or, equivalently \cite{Rovelli:2010qx}, as the limit for $\sigma\to\infty$ over the two-complexes $\sigma$ bounded by $\Gamma$, of the amplitude (we follow here \cite{Magliaro:2011qm} for the notation)
\begin{equation}\label{eq:GFTVE}
    \langle W_\sigma|\Gamma,j,n\rangle=\sum_{j_f}\int dg_{ve}\int d\vec n_{ef} \prod_f d_{j_{\!f}}\,   \tr\big[\prod_{e\in \p f} P_{ef}\big]
\end{equation}
where $e\in \p f$  is the ordered sequence of the oriented edges around the face $f$ and
\begin{equation}\label{eq:GFTVE}
P_{ef}=g_{s_ee}Y|j_f,\vec n_{ef}\rangle\langle j_f,-\vec n_{ef}|Y^\dagger g_{t_ee}^{-1}.
\end{equation}
for an internal edge $e$. For an external edge $e$, namely an edge hitting the boundary $\Gamma$ of $\sigma$,
\begin{equation}\label{eq:GFTVE}
P_{ef}=\langle j_l,-\vec n_{nl}|Y^\dagger g_{t_ee}^{-1}, \ \ \
{\rm or} \ \ \
P_{ef}=g_{s_ee}Y|j_l,\vec n_{nl}\rangle
\end{equation}
according to whether the orientation of the edge is incoming or outgoing. Here $l$ is the link bounding the face $f$ and $n$ is the node bounding the edge $e$. In all these formulas, the notation $g$ stands for the matrix elements of the group element $g$ in the appropriate  representation.

Here we deal with the Euclidean theory. Then $g_{ev}=(g^+_{ev},g^-_{ev})\in Spin(4)\sim SU(2)\times SU(2)$ and $Y$ maps the $SU(2)$ representations $j$ of  into the highest weight $SU(2)$ irreducible of the $SO(4)$ representation $(j^+,j^-)$, where $j^\pm=\frac12(1\pm \gamma)j$. The matrix elements of $Y$ are the standard Clebsch-Gordan coefficients.

The amplitude can be written in the form of a path integral by defining the action
\begin{equation}\label{eq:GFTVE}
S=\sum_f S_f= \sum_f \ln \tr\big[\prod_{e\in f} P_{ef}\big].
\end{equation}
Then
\begin{equation}\label{eq:GFTVE3}
     \langle W_\sigma|\Gamma,j,n\rangle=\sum_{j_f}\mu\int dg_{ve}\int d\vec n_{ef}\ e^S,
    \end{equation}
where $\mu=\sum_f d_j$. This is the form which is suitable for the asymptotic expansion that we use below.

Since the coherent states factorize under the Clebsch-Gordan decomposition, and since the scalar product of coherent states in the representation $j$ is the $j$'s power of that in the fundamental representation, we obtain $S=S^++S^-$ with \begin{equation}
S^{\pm}=\sum_{vf}2j^\pm_{f}\ln\bra{-\vec{n}_{ef}}(g_{ve}^{\pm})^{-1}g_{ve'}^{\pm}\ket{\vec{n}_{e'f}}
\end{equation}
where $e$ and $e'$ are the two edges bounding $f$ and $v$.

The last ingredient we need are the gravitational field operators $\gamma(x)$ that enter in Eq.\eqref{eq:QG}. The gravitational field operator that corresponds to the metric is expressed in loop quantum gravity by the Penrose operator \cite{Rovelli:2011eq}
 \begin{equation}
G_l^{ab}=E_l^a\cdot E_l^b,
 \end{equation}
 where $E_l^a$ is the left invariant vector field acting on the $h_{la}$ variable of the state vector, namely the $SU(2)$ group element associated to the link $a$ bounded by the node $l$.
The key technical observation of  \cite{Bianchi:2009ri} is that
 \begin{equation}\label{eq:OI}
    \left\langle W\left\vert G_l^{ab} \right\vert \Gamma ,j,n\right\rangle=
    \sum_{j_f}\mu \int dg_{ve}\int dn_{ef}\ q_l^{ab}\ \ex^S
  \end{equation}
where
$q_l^{ab}=A^{la}\cdot A^{lb}$, and $A_i^{la}=A_i^{la+}+A_i^{la-}$, \begin{equation}
    A_i^{la\pm}=\gamma j_{la}^{\pm}\frac{\bra{-\vec{n}_{al}}(g_a^{\pm})^{-1}g_l^{\pm}\sigma^i\ket{\vec{n}_{la}}}{\bra{-\vec{n}_{al}}(g_a^{\pm})^{-1}g_l^{\pm}\ket{\vec{n}_{la}}}.
  \end{equation}
This is the insertion that we consider below.

  \subsection{Vertex expansion}\label{sec:VA}

The second idea for computing $n$-point functions is the vertex expansion \cite{Rovelli:2005yj}.  This is the idea of studying the approximation to Eq.\eqref{eq:QG} given by the lowest order in the $\sigma\to\infty$ limit, namely using small graphs and small two-complexes.
Here we only look at the first nontrivial term. That is, we take a minimal two-complex, formed by a single vertex. We consider for simplicity the theory restricted to five-valent vertices and four-valent edges.  Then the lowest order is given by a two-complex formed by a single five-valent vertex bounded by the complete graph with 5  nodes $\Gamma_5$.  Labeling the nodes with  indices $a,b,...=1,...,5$ the amplitude of this two-complex for the boundary state $| \Gamma_5, j_{ab}, \vec n_{ab}\rangle$ (here $j_{ab}=j_{ba}$, but   $\vec n_{ab}\ne \vec n_{ba}$) reads simply
  \begin{equation}\label{eq:BDA}
    \langle W| \Gamma_5, j_{ab}, \vec n_{ab}\rangle=\mu(j)\int_{SU(2)^{10}} dg^\pm_a \  e^{\sum_{ab}S_{ab}}
  \end{equation}
  with
  \begin{equation}\label{eq:GFTVE6}
S_{ab}=\sum_{\pm}\ 2j_{ab}^\pm\ln\bra{-\vec{n}_{ab}}(g_{a}^{\pm})^{-1}g_{b}^{\pm}\ket{\vec{n}_{ba}}
 \end{equation}
The $\mu(j)$ term comes from the face amplitude and the measure (and cancels at the tree-level \cite{Bianchi:RC2008}).

The vertex expansion has appeared counterintuitive to some, on the base of the intuition  that the large distance limit of quantum gravity could be reached \emph{only} by states defined on very \emph{fine} graphs, and with very \emph{fine} two-complexes.  We are not persuaded by this intuition (in spite of the fact that one of the authors is quite responsible for propagandizing it \cite{Ashtekar:1992tm,Iwasaki:1992qy,Iwasaki:1994pb}) for a number of reasons.  The main one is the following.  It has been shown that under appropriate conditions Eq.\eqref{eq:GFTVE3} can approximates a Regge path integral for large spins \cite{Conrady:2008ea,Barrett:2009cj,Magliaro:2011qm}.  Regge calculus is an approximation to general relativity that is good up to order $\mathcal{O}(l^2/\rho^2)$, where $l$ is the typical Regge discretization length and $\rho$ is the typical curvature radius.  This implies that Regge theory on a coarse lattice is good as long as we look at small curvatures scale.  In particular, it is obviously perfectly good on flat space, where in fact it is exact, because the Regge simplices are themselves flat, and is good as long as we look at weak field perturbations of long wavelength. This is precisely the limit in which we want to study the theory here.  In this limit, it is therefore reasonable to explore whether the vertex expansion give any sensible result.

Reducing the theory to a single vertex is a drastic simplification of the field theory, which reduce the calculation to one for a system with a finite number of degrees of freedom. Is this reasonable? The answer is in noticing that the same drastic simplification occurs in  the analog calculation in QED: at the lowest order, an $n$-point function involves only the Hilbert space of a finite number of particles, which are described by a finite number of degrees of freedom in the classical theory. The genuine field theoretical aspects of the problem, such as renormalization, do not show up at the lowest order, of course.

If we regard the calculation from the perspective of the triangulation dual to the two-complex, what is being considered is a region of spacetime with the geometry of a 4-simplex. In the approximation considered the region is flat, but this does not mean that there are no degrees of freedom. In fact, the Hamilton function of general relativity is a nontrivial function of the intrinsic geometry of the boundary, whose variation gives equations that determine the extrinsic geometry as a function of the intrinsic geometry.  This relation captures a small finite-dimensional sector of the Einstein-equations dynamics (for a simple example of this, see \cite{Colosi:2004sa}).  This is precisely the component of the dynamics of general relativity captured in this limit.  The three-point function in this large wavelength limit describes the correlations between the fluctuations of the boundary geometry of the 4-simplex, governed by the quantum version of this restricted Einstein dynamics.

Let us illustrate this dynamics a bit more in detail, both in second order (metric) and first order (tetrad/connection) variables. In metric variables, the intrinsic geometry of a boundary of a four-simplex (formed by glued flat tetrahedra) is uniquely determined by the 10 areas $A_{ab}$ of their faces. The extrinsic geometry of the boundary four-simplex is determined by the 10 angles $\Phi_{ab}$ between the 4d normals to the tetrahedra.  The Einstein equations reduce in the case of a single simplex to the requirement that this is flat. If the four simplex is flat, then the 10 angles $\Phi_{ab}$ are well-defined functions
\begin{equation}\label{eq:angles}
\Phi_{ab}=\Phi_{ab}(A_{ab})
\end{equation}
 of the 10 areas $A_{ab}$ (for comparison, if the four-simplex has constant curvature because of a cosmological constant, then the same $A_{ab}$'s determine different
$\Phi_{ab}$'s). This dependence captures the restriction of the Einstein equations to a single simplex. In first order variables, the situation is more complicated. The variables $g$, $j$ and $\vec n$ in Eq.\eqref{eq:GFTVE} can be viewed as the discretized version of the connection and the tetrad. The vanishing-torsion equation of the first order formalism, which relates the connection to the tetrad, becomes in the discrete formalism a gluing condition between normals to the faces parallel transported by the group elements.

\subsection{Boundary vacuum state}\label{sec:Bn}

Following the general strategy described above, we need a boundary state peaked on the intrinsic as well as on the extrinsic geometry. This state cannot be the state  $| \Gamma_5, j_{ab}, \vec n_{ab}\rangle$ which is an eigenvalue of boundary areas, and therefore is maximally spread in the  extrinsic curvature, namely in the 4d dihedral angle between two boundary tetrahedra $\Phi_{ab}$ \cite{Rovelli:2010km}. Rather, we need a state which is also smeared over spins \cite{Sahlmann:2001nv,Thiemann:2002vj,Bianchi:2009ky}.

Following \cite{Bianchi:2009ky}, we choose here a boundary state peaked on the intrinsic and extrinsic geometry of a \emph{regular} 4-simplex, and defined as follow.  The geometry of a \emph{flat} 4-simplex is uniquely determined by the 10 areas $A_{ab}$ of its 10 faces. Let then $\vec n_{ab}(A_{ab})$ be the 20 normals determined up to arbitrary $SO(3)$ rotations of each quadruplet $\vec n_{ab_1},...,\vec n_{ab_4}$ by these areas. By this we mean the following. The flat 4-simplex determined by the given areas is bounded by five tetrahedra. For each such tetrahedron, the four normals to its four faces in the 3-space determined by the tetrahedron determine, up to rotations) the four unit vectors $\vec n_{ab_1},...,\vec n_{ab_4}$.  Using this, we define the boundary state as
  \begin{equation}\label{eq:VS}
    \ket{\Psi_{\Sigma}}= \ket{\Psi_{j_0}}=\sum_{j_{ab}} c_{{j_0}}(j_{ab})\ket{\Gamma,j_{ab},n_{ab}(j_{ab})}
  \end{equation}
  where the coefficients $c_{{j_0}}(j)$ in the large $j$ limit are given by \cite{Bianchi:2009ky}
  \begin{equation}\label{eq:VSc}
    c_{{j_0}}(j_{ab})=\frac{1}{N}e^{-\sum_{(ab),(cd)}\gamma\alpha^{(ab)(cd)}\frac{j_{ab}-j_{0}}{\sqrt{j_{0}}}\frac{j_{cd}-j_{0}}{\sqrt{j_{0}}}-\im \sum_{(ab)}\Phi_{0}\gamma j_{ab}}
  \end{equation}
  The coefficients are also given in \cite{Rovelli:2005yj,Bianchi:2006uf}. $\alpha^{(ab)(cd)}$ is a $10\times10$ matrix that has the symmetries of the 4-simplex, that is, it can be written in the form $\alpha^{(ab)(cd)}=\sum_k\alpha_kP_k^{(ab)(cd)}$ where
  \begin{align}
    P_0^{(ab)(cd)}=1 &\quad \text{if}\quad (ab)=(cd)\quad \text{and 0 otherwise},\nonumber\\
    P_1^{(ab)(cd)}=1 &\quad \text{if}\quad  \{a=c,b\neq d\} \;\;\text{or a permutation},\nonumber \\
    &\hspace{6em} \text{and 0 otherwise},\nonumber\\
    P_2^{(ab)(cd)}=1 &\quad \text{if}\quad (ab)\neq(cd)\quad \text{and 0 otherwise}.\nonumber
  \end{align}
$\Phi_{0}$ is the background value of the 4d dihedral angles which give the extrinsic curvature of the boundary.  $j_0$ is the background value of all the areas. The state is peaked on the areas $j_{ab}=j_0$, which determine a regular 4-simplex. The dihedral angles of a flat tetrahedron is $\Phi_{0}= \arccos(-\frac14)$, and we fix $\Phi_{0}$ to this value. As a consequence $\ket{\Psi_{j_0}}$ is a semiclassical \emph{physical} state, namely it is peaked on values of intrinsic and extrinsic geometry that satisfy the (Hamilton) equations of motion \eqref{eq:angles} of the theory.
See \cite{Bianchi:2006uf,Bianchi:2009ri,Rovelli:2005yj,Colosi:2004sa} for more details.

  \subsection{Three-point function}\label{sec:E}

Let us now choose the operator insertion.  We are interested in the connected component of the quantity
 \begin{equation}\label{eq:def}
      \tilde G_{lmn}^{abcdef}=\langle G_l^{ab}~G_m^{cd}~G_n^{ef}\rangle,
        \end{equation}
where $G_l^{ab}$ is the Penrose operator associated to the node $l$ of $\Gamma_5$ and the two links of this node going from $l$ to $a$ and from $l$ to $b$ respectively.
 The connected component is
 \begin{equation}\label{eq:defc}
    \begin{split}
      G_{lmn}^{abcdef}&=\langle G_l^{ab}~G_m^{cd}~G_n^{ef}\rangle +2\langle G_l^{ab}\rangle\langle G_m^{cd}\rangle\langle G_n^{ef}\rangle \\
      &\quad-\langle G_l^{ab}\rangle\langle G_m^{cd}~G_n^{ef}\rangle
      -\langle G_n^{ef}\rangle\langle G_l^{ab} G_m^{cd}\rangle
      \\ &\quad
      -\langle G_m^{cd}\rangle\langle G_l^{ab} G_n^{ef}\rangle
    \end{split}
  \end{equation}
We begin by studying the full three-point function Eq.\eqref{eq:def}, before subtracting the disconnected components.  From Eq.\eqref{eq:QG} and Eq.(\ref{eq:VSc}), and simplifying a bit the notation in a self explicatory way, this is
  \begin{equation}
    \begin{split}
      \tilde G_{lmn}^{abcdef} &=\frac{\sum_{j}c\left( j\right) \left\langle W\left\vert
G_l^{ab}\,G_m^{cd}\,G_n^{ef}\right\vert \Gamma_5 ,j,n\right\rangle }{\sum_{j}c\left( j\right)
\left\langle W|\Gamma ,j,n\right\rangle }    \end{split}
  \end{equation}
Using Eq.\eqref{eq:OI}, this gives
  \begin{equation}
    \begin{split}
     \tilde G_{lmn}^{abcdef} &=\frac{\sum_{j}c\left( j\right) \int dg_a^\pm
q_l^{ab}\,q_m^{cd}\,q_n^{ef} e^S }{\sum_{j}c\left( j\right) \int dg_a^\pm
 e^S }    \end{split}
  \end{equation}
where the sum over spins is only given by the boundary state, since there are no internal faces.

Define the total action as $S_{\text{tot}}=\ln c(j)+S$. Because we want to get the large $j$ limit of the spin foam model, we rescale the spins $j_{ab}$ and $j_{0}$. Then the action goes to $S_{\text{tot}}\rightarrow\lambda S_{\text{tot}}$ and also $q_l^{ab}\rightarrow\lambda^2q_n^{ab}$. In large $\lambda$ limit, the sum over $j$ can be approximated to the integrals over $j$
  \begin{equation}
    \sum_j\mu\int\dd g_a^{\pm}\, q_l^{ab}\, \ex^{\lambda S_{\text{tot}}}\approx\int\dd j\dd g_a^{\pm}\mu \, q_l^{ab}\, \ex^{\lambda S_{\text{tot}}}
  \end{equation}
  where $\mu$ is the product of the face amplitudes.
  Thus (dropping the suffix $tot$ from now on)
    \begin{equation}\label{eq:def2}
    \begin{split}
  &  \tilde G_{lmn}^{abcdef}=\lambda^6\ \ \dfrac{\int\dd j\dd g_a^{\pm}\mu q_l^{ab}q_m^{cd}q_n^{ef}\ex^{\lambda S}}{\int\dd j\dd g_a^{\pm}\mu\ex^{\lambda S}}\\
    \end{split}
  \end{equation}
 Action, measure and insertions are invariant under a $SO(4)$ symmetry, therefore only four of the five $\dd g_a^\pm$ integrals are independent. We can fix the gauge that one of the group element $g^{\pm}=\mathds{1}$, and the integral reduced to $\dd g=\prod_{a=1}^4\dd g^+_a\dd g^-_a$. This gives the expression
     \begin{equation}\label{eq:def2}
    \begin{split}
    \tilde G_{lmn}^{abcdef}&=\lambda^6\ \ \dfrac{\int\dd j\dd  g\ \mu q_l^{ab}q_m^{cd}q_n^{ef}\ex^{\lambda S}}{\int\dd j\dd  g\ \mu\ex^{\lambda S}}
     \end{split}
  \end{equation}
We simplify the notation by writing this in the simple form
     \begin{equation}\label{eq:def2}
    \begin{split}
    \tilde G&=\lambda^6\ \ \dfrac{\int\dd j\dd  g\ \mu\, l\, m\, n\, \ex^{\lambda S}}{\int\dd j\dd  g\ \mu\ex^{\lambda S}} \equiv  \langle lmn \rangle
     \end{split}
  \end{equation}
where $l=q_l^{ab},m=q_m^{cd},n=q_n^{ef}$ are functions of $j$ and $g$.
The connected component reads then
  \begin{equation}\label{eq:def22}
      G=\langle lmn \rangle
      +2\langle l \rangle\langle  m \rangle \langle n \rangle
      - \langle lm \rangle \langle n\rangle
         -\langle nl \rangle\langle m \rangle
            - \langle  mn \rangle\langle l \rangle
  \end{equation}
 which is the point of departure for the saddle point expansion.

  \subsection{Saddle point expansion}
  To study the asymptotic behavior of Eq.(\ref{eq:def2}), we use the saddle point expansion\cite{Zee:QFT,saddle,Bianchi:RC2008}. For this, we need the stationary point of the total action $S_{\text{tot}}=\ln c(j)+S$. Here we briefly review the works in \cite{Barrett:2009gg} and \cite{Bianchi:2009ri}. They discuss the behavior of the critical point and stationary point of $S=S^++S^-$ and $S_{\text{tot}}$. We invite readers to read their articles for full detail discussion.

  The critical point and stationary point of $\Rl S$ coincide with each other when $\gamma<1$. For the real part of the action $S$, the critical points are the group element $\bar{g}^{\pm}$ satisfy the gluing condition
  \begin{equation}\label{eq:glu}
    R_a^{\pm}n_{ab}=-R_b^{\pm}n_{ba}
  \end{equation}
where $R_a^{\pm}=R(\bar{g}_a^{\pm})$ is the spin-$1$ irrep.\,of SU(2). This means that at the critical point the geometry of spacetime goes to a classical one in which all tetrahedra glue perfectly. There are 4 classes of critical points satisfy the condition (\ref{eq:glu}). At the critical points of $\Rl S$, the action $S$ can be written as $S=\im A$, where $A$ is a real function and  reduces to Regge like actions. See \cite{Barrett:2009gg} and \cite{Bianchi:2009ri}. A unique class of critical points is then selected by the stationary point behavior of $S_{\text{tot}}$.

The stationary points of $\Rl S$ are the critical points of $\Rl S$, because of the closure constraint, which is satisfied by the boundary state for large $j_0$. We are interested in the  stationary points of $S_{\text{tot}}$ are not just with respect to the group variables, but also with respect to the spin $j$ variables. The stationary point $j_{ab}=j_0$ also selects the class of group stationary point. This is because at the stationary point, $S$ must satisfy
  \begin{equation}
    -\im \gamma\Phi_{ab}+\frac{\p S(g_0)}{\p j_{ab}}=0
  \end{equation}
Therefore it means that only when $S(g_0)=\im S_{\text{Regge}}$ (with a definite sign) this condition can be satisfied. This condition picks the unique class of critical points $g_0^{\pm}$ of $\Rl S$, which makes $S(g_0)=\im S_{\text{Regge}}$.

We are thus interested in the saddle point expansion of the integrals in Eq.(\ref{eq:def2}) around the stationary points ($j_0,g_0^{\pm}$) described above. According to the general theory, the integral
  \begin{equation}\label{eq:integral}
    F(\lambda)=\int\dd xf(x)\ex^{\lambda S(x)}
  \end{equation}
can be expand for large $\lambda$ around the stationary points as follows
  \begin{eqnarray}\label{eq:defF}
    F(\lambda)&=&C(x_0)\left(f(x_0)+\frac{1}{\lambda}\left(\frac{1}{2}f_{ij}(x_0)J^{ij}+D\right)\right)\nonumber \\ && +\mathcal{O}\left(\frac{1}{\lambda^2}\right)
  \end{eqnarray}
where $x_0$ is the stationary point, $f_{ij}$ is the Jacobian matrix of $f$,  and $J=H^{-1}=(S^{\prime\prime}(x_0))^{-1}$ is the inverse of the Jacobian matrix of the action $S$.  A straightforward application of this formula to Eq.\eqref{eq:def22} shows that
  \begin{equation}
    G=0+\mathcal{O}\left(\frac{1}{\lambda^2}\right)
  \end{equation}
 This in fact is not surprising, because we are computing a three point function, and this cannot be captured only by the second order of the saddle point expansion. The second order of the saddle point expansion sees only the second derivatives of the action, while the connected component of the three point function depends on the third derivatives of the action.  In fact, the 3rd derivative of the action term can be identified with a Feynman vertex, the inverse of the second derivative as the propagator and the insertions as the external legs of a Feynman diagram. Then It is clear that to second order there is no connected component.

 Therefore we need the next order of the saddle point expansion.
From  Eq.(\ref{eq:integral}), this is given by
  \begin{eqnarray}\label{eq:defF}
    F(\lambda)&=&C(x_0)\left(f(x_0)+\frac{F_1}{\lambda}+\frac{F_2}{\lambda^2}\right)
    +\mathcal{O}\left(\frac{1}{\lambda^3}\right)
  \end{eqnarray}
  where
  \begin{eqnarray}\label{lambda1}
     F_1&=&-\frac{1}{2}f_{ij}J^{ij}+\frac{1}{2}f_i J^{ij}J^{kl}R_{jkl}\\
      &&\quad-\frac{5}{24}fJ^{il}J^{jm}J^{kn}R_{ijk}R_{lmn}
      +\frac{1}{8}fJ^{ik}J^{jl}R_{ijkl}
      \nonumber
  \end{eqnarray}
  and
  \begin{equation}\label{lambda2}
    \begin{split}
      F_2&=\frac{1}{8}f_{ijkl}J^{ij}J^{kl}-\frac{5}{12}f_{ijk}J^{il}J^{jm}J^{kn}R_{lmn}
      \\&\quad
      -\frac{5}{16}f_{ij}J^{ik}J^{jl}J^{mn}R_{klmn}
     \\&\quad
       +\frac{35}{48}f_{ij}J^{im}J^{jn}J^{ko}J^{lp}R_{mko}R_{nlp}
       +\cdots
    \end{split}
  \end{equation}
 Here $R(x)=S(x)-S-\dfrac{1}{2}H_{ij}(x-x_0)^i(x-\bar{x})^j$, all functions are computed in $x_0$,  the stationary point of $S(x)$ and the indices indicate derivatives. In the last two equations we have left understood some symmetrization. For instance the third term of the right hand side of Eq.(\ref{lambda1}) should read .
  \begin{equation}
  \frac{5}{48}f\left(J^{il}J^{jm}J^{kn}+J^{il}J^{jm}J^{kn}\right)R_{ijk}R_{lmn}
  \end{equation}
and so on.

Using this, and recalling that here $f(x)=\mu(x) l(x) m(x) n(x)$, we obtain, up to order $\mathcal{O}(\frac{1}{\lambda^2})$,
  \begin{eqnarray}\label{eq:3pS}
G_{lmn}^{abcdef}\!\!&=\!\!&\lambda^4\Big(\!-R_{ijk} l _l m _m n _n J^{il}J^{jm}J^{kn}\\\nonumber
      &&+(l_{ij} m _k n _l
      +l_k m_{ij} n _l
       +l _k m _l n_{ij} )J^{ik}J^{jl}\Big)
  \end{eqnarray}
The first term on the right hand side corresponds to the one vertex diagram with three legs. The second term corresponds to the 4-point function in which 2 points are identified. In perturbation theory, this term is divergent. We will see that this term is finite here.

  \subsection{Analytical expression}\label{sec:AE}
Eq.(\ref{eq:3pS}) indicates that we need to get the second and  third derivatives of the total action, and the first and second derivative of the insertions. Here we compute these terms.

We use Euler angles to parameterize the $SU(2)$ group elements $g_0^{\pm}$ around the stationary point
  \begin{equation}
    R_a^{\pm}=\ex^{\im\theta_i^{\pm}J_i}R_{0a}^{\pm}
  \end{equation}
  where $i=1,2,3$, $\theta_i$ are Euler angles, $J_i$ are the generators of SU(2), $R_a$ stands for arbitrary irrep. of SU(2). There are 34 independent variables, 10 areas $j_{ab}$ of triangles in the 4 simplex, 24 group element parameters in which 12 for $g^+$ and 12 for $g^-$. Here we give only some steps to get to the result. The whole results can be found in the Appendix.
The second order derivative of the total action gives
    \begin{eqnarray}
      \frac{\p^2 S_{\text{tot}}}{\p j_{ab}\p j_{cd}}\bigg|_{\theta=0}\hspace{-1em}&=&-\frac{\gamma\alpha^{(ab)(cd)}}{\sqrt{j_{0ab}}\sqrt{j_{0cd}}}+\im\frac{\p^2 S_{\text{Regge}}}{\p j_{ab}\p j_{cd}}\\
      \frac{\partial ^{2}S}{\partial \theta _{j}^{a\pm }\partial \theta _{i}^{a\pm
}}\bigg|_{\theta =0}\hspace{-1em}&=&-\frac{\gamma ^{\pm }}2\sum_{\left( b\neq a\right)
}j_{ab}\left( \delta _{ij}-\left( n_{ab}^{\pm }\right) _{i}\left(
n_{ab}^{\pm }\right) _{j}\right)\\
\frac{\partial ^{2}S}{\partial \theta _{j}^{b\pm }\partial \theta _{i}^{a\pm
}}\bigg|_{\theta =0}\hspace{-1em}&=&\frac{\gamma ^{\pm }j_{ab}}2(\delta _{ij}\!-\!\left(
n_{ab}^{\pm }\right) _{i}\! \left( n_{ab}^{\pm }\right) _{j}\!-\! {i\varepsilon
_{ijk}}\!\left( n_{ab}^{\pm }\right) _{k})\nonumber \\
    \end{eqnarray}
 The third order derivative of the total action gives
\begin{widetext}
    \begin{eqnarray}
      \frac{\p^3 S_{\text{tot}}}{\p j_{ab}\p j_{cd}\p j_{ef}}\bigg|_{\theta=0}&=&\im\frac{\p^3 S_{\text{Regge}}}{\p j_{ab}\p j_{cd}\p j_{ef}}\\
      \frac{\partial ^{3}S}{\partial \theta _{k}^{a\pm }\partial \theta _{j}^{a\pm
}\partial \theta _{i}^{a\pm }}\bigg|_{\theta =0}&=&\sum_{b\neq a}\frac{1}{6}i\gamma
^{\pm }j_{ab}(\delta _{jk}\left( n_{ab}^{\pm }\right) _{i}+\delta
_{ki}\left( n_{ab}^{\pm }\right) _{j}+\allowbreak \delta _{ij}\left(
n_{ab}^{\pm }\right) _{k}-3\left( n_{ab}^{\pm }\right) _{i}\left(
n_{ab}^{\pm }\right) _{j}\left( n_{ab}^{\pm }\right) _{k})\\
\frac{\partial ^{3}S}{\partial \theta _{k}^{b\pm }\partial \theta _{j}^{a\pm
}\partial \theta _{i}^{a\pm }}|_{\theta =0} &=&-\frac{1}{4}i\gamma ^{\pm }j_{ab}(\delta _{jk}\left( n_{ab}^{\pm }\right)
_{i}+\delta _{ki}\left( n_{ab}^{\pm }\right) _{j}-2\left( n_{ab}^{\pm
}\right) _{i}\left( n_{ab}^{\pm }\right) _{j}\left( n_{ab}^{\pm }\right)
_{k}\\ \nonumber
&&+i\left( \varepsilon _{kil}\left( n_{ab}^{\pm }\right) _{j}+\varepsilon
_{kjl}\left( n_{ab}^{\pm }\right) _{i}\right) \left( n_{ab}^{\pm }\right)
_{l})\\
\cdots
    \end{eqnarray}
     The first derivatives of the insertions
        \begin{eqnarray}
          \frac{\partial q_{c}^{ab}}{\partial j_{ef}} &=&\gamma ^{2}\frac{\partial
(j_{ca}j_{cb}n_{ca}\cdot n_{cb})}{\partial j_{ef}}=\gamma ^{2}\frac{\partial
(j_{ca}j_{cb}\cos \Theta _{cab})}{\partial j_{ef}} \\
\frac{\partial q_{n}^{ab}}{\partial \theta _{i}^{a\pm }}|_{\theta _{i}^{a\pm
}=0}&=&-\frac{1}{2}i\gamma ^{2}\gamma ^{\pm }j_{na}j_{nb}(\left( n_{nb}^{\pm
}\right) _{i}-\left( n_{na}^{\pm }\right) _{i}\left( n_{nb}\right)
_{j}\left( n_{na}\right) _{j}+i\varepsilon _{ijk}\left( n_{nb}^{\pm }\right)
_{j}\left( n_{na}^{\pm }\right) _{k})\\
\frac{\partial q_{n}^{ab}}{\partial \theta _{i}^{n\pm }}|_{\theta _{i}^{n\pm
}=0}&=&\frac{1}{2}i\gamma ^{2}\gamma ^{\pm }j_{na}j_{nb}\left( \left(
n_{na}^{\pm }\right) _{i}+\left( n_{nb}^{\pm }\right) _{i}\right) (1-\left(
n_{na}\right) _{j}\left( n_{nb}\right) _{j})
        \end{eqnarray}
         The second derivatives of the insertions
            \begin{eqnarray}
              \frac{\partial ^{2}q_{n}^{ab}}{\partial \theta _{j}^{a\pm }\partial \theta
_{i}^{a\pm }}|_{\theta =0}&=&\frac{1}{4}\gamma ^{2}\gamma ^{\pm }j_{na}j_{nb}(\left( n_{nb}^{\pm
}\right) _{j}\left( n_{na}^{\pm }\right) _{i}+\left( n_{nb}^{\pm }\right)
_{i}\left( n_{na}^{\pm }\right) _{j}-2\left( n_{nb}\right) _{r}\left(
n_{na}\right) _{r}\left( n_{na}^{\pm }\right) _{i}\left( n_{na}^{\pm
}\right) _{j} \\ \nonumber
&&-i\left( n_{nb}^{\pm }\right) _{k}\left( n_{na}^{\pm }\right) _{m}\left(
\varepsilon _{kmj}\left( n_{na}^{\pm }\right) _{i}+\varepsilon _{kmi}\left(
n_{na}^{\pm }\right) _{j}\right) )\\
\cdots
            \end{eqnarray}
 \end{widetext}

  \subsection{Numerical results}
  The derivatives over the spin $j$s can be obtained numerically. For simplicity, we only consider the situation where the boundary is a regular 4-simplex.
  For the total action $S$, the second derivatives

\begin{eqnarray*}
\frac{\partial ^{2}S}{\partial j_{ab}\partial j_{ab}} &=&-\frac{\gamma\alpha_0}{j_0}-%
\im\frac{\gamma }{j_{0}}\frac{9}{4}\sqrt{\frac{3}{5}} \\
\frac{\partial ^{2}S}{\partial j_{ac}\partial j_{ab}} &=&-\frac{\gamma\alpha_1}{j_0}+%
\im\frac{\gamma }{j_{0}}\frac{8}{7}\sqrt{\frac{3}{5}} \\
\frac{\partial ^{2}S}{\partial j_{cd}\partial j_{ab}} &=&-\frac{\gamma\alpha_2}{j_0}-%
\im\frac{\gamma }{j_{0}}\sqrt{\frac{3}{5}}
\end{eqnarray*}

For the third derivatives, only seven of them are independent. They are
\begin{eqnarray*}
\frac{\partial ^{3}S}{\partial j_{ab}\partial j_{ab}\partial j_{ab}} &=&-i%
\frac{\gamma }{j_{0}^{2}}\frac{189}{80}\sqrt{\frac{3}{5}},
\\ \quad\frac{\partial
^{3}S}{\partial j_{ac}\partial j_{ab}\partial j_{ab}}&=&i\frac{\gamma }{%
j_{0}^{2}}\frac{347}{160}\sqrt{\frac{3}{5}}  \\ \quad
\frac{\partial ^{3}S}{\partial j_{cd}\partial j_{ab}\partial j_{ab}} &=&-i%
\frac{\gamma }{j_{0}^{2}}\frac{14}{5}\sqrt{\frac{3}{5}},\\
\quad\frac{\partial ^{3}S%
}{\partial j_{ad}\partial j_{ac}\partial j_{ab}}&=&-i\frac{\gamma }{j_{0}^{2}}%
\frac{453}{160}\sqrt{\frac{3}{5}} \\
\frac{\partial ^{3}S}{\partial j_{bc}\partial j_{ac}\partial j_{ab}} &=&-i%
\frac{\gamma }{j_{0}^{2}}\frac{141}{20}\sqrt{\frac{3}{5}},\\
\quad\frac{\partial
^{3}S}{\partial j_{bd}\partial j_{ac}\partial j_{ab}}&=&i\frac{\gamma }{%
j_{0}^{2}}\frac{39}{20}\sqrt{\frac{3}{5}} \\
\frac{\partial ^{3}S}{\partial j_{ed}\partial j_{ac}\partial j_{ab}} &=&-i%
\frac{\gamma }{j_{0}^{2}}\frac{3}{10}\sqrt{\frac{3}{5}}
\end{eqnarray*}

For the metric quantities $q_{n}^{ab}$, when $a\neq b$, we can find there
are only five of them are independent. They are%
\begin{eqnarray*}
\frac{\partial q_{c}^{ab}}{\partial j_{ab}} &=&\frac{4}{3}\gamma ^{2}j_{0},\quad%
\frac{\partial q_{c}^{ab}}{\partial j_{ac}}=-\frac{2}{3}\gamma ^{2}j_{0} \\
\frac{\partial q_{c}^{ab}}{\partial j_{ad}} &=&-\frac{2}{3}\gamma ^{2}j_{0},\quad%
\frac{\partial q_{c}^{ab}}{\partial j_{cd}}=\frac{1}{3}\gamma ^{2}j_{0} \\
\frac{\partial q_{c}^{ab}}{\partial j_{de}} &=&\frac{4}{3}\gamma ^{2}j_{0}
\end{eqnarray*}

The second derivatives%
\begin{eqnarray*}
\frac{\partial ^{2}q_{c}^{ab}}{\partial j_{gh}\partial j_{ef}} &=&\gamma ^{2}%
\frac{\partial ^{2}(j_{ca}j_{cb}n_{ca}\cdot n_{cb})}{\partial j_{gh}\partial
j_{ef}} \\
&&\hspace{-5em}=\gamma^{2}{\scriptscriptstyle \left(
\begin{array}{cccccccccc}
\frac{4}{3} & 1 & -2 & -2 & 1 & -2 & -2 & 1 & 1 & 4 \\
1 & -\frac{2}{3} & -\frac{1}{2} & -\frac{1}{2} & 1 & -\frac{1}{2} & -\frac{1%
}{2} & -\frac{1}{2} & -\frac{1}{2} & 1 \\
-2 & -\frac{1}{2} & -\frac{2}{3} & 4 & -\frac{1}{2} & 4 & -2 & -\frac{1}{2}
& -\frac{1}{2} & -2 \\
-2 & -\frac{1}{2} & 4 & -\frac{2}{3} & -\frac{1}{2} & -2 & 4 & -\frac{1}{2}
& -\frac{1}{2} & -2 \\
1 & 1 & -\frac{1}{2} & -\frac{1}{2} & -\frac{2}{3} & -\frac{1}{2} & -\frac{1%
}{2} & -\frac{1}{2} & -\frac{1}{2} & 1 \\
-2 & -\frac{1}{2} & 4 & -2 & -\frac{1}{2} & -\frac{2}{3} & 4 & -\frac{1}{2}
& -\frac{1}{2} & -2 \\
-2 & -\frac{1}{2} & -2 & 4 & -\frac{1}{2} & 4 & -\frac{2}{3} & -\frac{1}{2}
& -\frac{1}{2} & -2 \\
1 & -\frac{1}{2} & -\frac{1}{2} & -\frac{1}{2} & -\frac{1}{2} & -\frac{1}{2}
& -\frac{1}{2} & \frac{1}{3} & -1 & 1 \\
1 & -\frac{1}{2} & -\frac{1}{2} & -\frac{1}{2} & -\frac{1}{2} & -\frac{1}{2}
& -\frac{1}{2} & 1 & \frac{1}{3} & 1 \\
4 & 1 & -2 & -2 & 1 & -2 & -2 & 1 & 1 & \frac{4}{3}%
\end{array}%
\right)}
\end{eqnarray*}
Rows and columns are labeled by $j_{gh}$ and $j_{ef}$, respectively. The
order is $%
\{j_{ab},j_{ac},j_{ad}, j_{ae},j_{bc},j_{bd},j_{be},j_{cd},j_{ce},  $ $ j_{de}\}$.

When $a=b$, there is only one non-zero first and second
derivatives. They are%
\begin{eqnarray*}
\frac{\partial q_{c}^{aa}}{\partial j_{ca}} &=&2\gamma ^{2}j_{0}, \ \ \ \
\frac{\partial ^{2}q_{c}^{aa}}{\partial j_{ca}\partial j_{ca}} =2\gamma
^{2}
\end{eqnarray*}

Now, let us look at the dependence of these quantities from $\gamma$ and $j=j_0$. We obtain
\begin{eqnarray*}
\frac{\partial ^{3}S}{\partial j\partial j\partial j} &\sim &\frac{\gamma }{%
j^{2}},\,\quad \quad \frac{\partial ^{2}S}{\partial j\partial j}\sim \frac{%
\gamma }{j},\,\quad \quad \frac{\partial ^{3}S}{\partial \theta \partial
\theta \partial \theta }\sim \gamma ^{\pm }j, \\  \quad \frac{\partial ^{2}S%
}{\partial \theta \partial \theta }&\sim & \gamma ^{\pm }j,\,\quad  \frac{%
\partial ^{3}S}{\partial j\partial \theta \partial \theta }\sim \gamma ^{\pm
},  \quad \quad
\frac{\partial ^{2}q}{\partial j\partial j} \sim \gamma ^{2},\\
\,\quad
\frac{\partial q}{\partial j}&\sim & \gamma ^{2}j,\,\quad \quad \frac{\partial
^{2}q}{\partial \theta \partial \theta }\sim \gamma ^{2}\gamma ^{\pm
}j^{2},\quad \quad \frac{\partial q}{\partial \theta }\sim \gamma ^{2}\gamma
^{\pm }j^{2}, \\  \frac{\partial ^{2}q}{\partial j\partial \theta }%
&\sim& \gamma ^{2}\gamma ^{\pm }j.
\end{eqnarray*}
For the 3-valent term, the scaling is
  \begin{equation}\label{resta}
\frac{\partial ^{3}S}{\partial j\partial j\partial j}\left(\! \frac{\partial
^{2}S}{\partial j\partial j}\!\right) ^{\!\!\!-1}\!\!\left(\! \frac{\partial ^{2}S}{%
\partial j\partial j}\!\right) ^{\!\!\!-1}\!\!\left(\! \frac{\partial ^{2}S}{\partial
j\partial j}\!\right) ^{\!\!\!-1}\!\frac{\partial q}{\partial j}\frac{\partial q}{%
\partial j}\frac{\partial q}{\partial j} \sim \gamma ^{4}j^{4}
\end{equation}
and
\begin{eqnarray*}
&&\frac{\partial ^{3}S}{\partial \theta \partial \theta \partial \theta }%
\left(\! \frac{\partial ^{2}S}{\partial \theta \partial \theta }\!\right)
^{\!\!-1}\!\left(\! \frac{\partial ^{2}S}{\partial \theta \partial \theta }\!\right)
^{\!\!-1}\!\left(\! \frac{\partial ^{2}S}{\partial \theta \partial \theta }\!\right)
^{\!\!-1}\!\frac{\partial q}{\partial \theta }\frac{\partial q}{\partial \theta }%
\frac{\partial q}{\partial \theta }
\\ &&\hspace{5em}\sim \gamma ^{\pm }\gamma
^{6}j^{4}\rightarrow \gamma ^{6}j^{4}|_{\gamma \rightarrow 0} \\
&&\frac{\partial ^{3}S}{\partial j\partial \theta \partial \theta }\left(\!
\frac{\partial ^{2}S}{\partial \theta \partial \theta }\!\right) ^{\!\!-1}\!\left(\!
\frac{\partial ^{2}S}{\partial \theta \partial \theta }\!\right) ^{\!\!-1}\!\left(\!
\frac{\partial ^{2}S}{\partial j\partial j}\!\right) ^{\!\!-1}\!\frac{\partial q}{%
\partial \theta }\frac{\partial q}{\partial \theta }\frac{\partial q}{%
\partial j}
\\  &&\hspace{5em} \sim \gamma ^{\pm }\gamma ^{5}j^{4}\rightarrow \gamma
^{5}j^{4}|_{\gamma \rightarrow 0}
\end{eqnarray*}%
  And for the ``4''-point function terms,
  \begin{equation}\label{resta2}
\frac{\partial ^{2}q}{\partial j\partial j}\left(\!\frac{\partial ^{2}S}{%
\partial j\partial j}\right)^{\!\!\!-1}\!\left(\! \frac{\partial ^{2}S}{\partial
j\partial j}\!\right)^{\!\!\!-1}\frac{\partial q}{\partial j}\frac{\partial q}{%
\partial j} \sim \gamma ^{4}j^{4}
  \end{equation}
and
  \begin{eqnarray*}
\frac{\partial ^{2}q}{\partial \theta \partial \theta }\left(\! \frac{\partial
^{2}S}{\partial \theta \partial \theta }\!\right)^{\!\!\!-1}\!\left(\! \frac{\partial
^{2}S}{\partial \theta \partial \theta }\!\right) ^{\!\!\!-1}\frac{\partial q}{%
\partial \theta }\frac{\partial q}{\partial \theta } &\sim\! &\gamma ^{\pm
}\gamma ^{6}j^{4}\rightarrow \gamma ^{6}j^{4}|_{\gamma \rightarrow 0} \\
\frac{\partial ^{2}q}{\partial j\partial \theta }\left(\! \frac{\partial ^{2}S%
}{\partial \theta \partial \theta }\!\right)^{\!\!\!-1}\!\left(\! \frac{\partial ^{2}S}{%
\partial j\partial j}\!\right) ^{\!\!\!-1}\frac{\partial q}{\partial \theta }\frac{%
\partial q}{\partial j} &\sim\! &\gamma ^{\pm }\gamma ^{5}j^{4}\rightarrow
\gamma ^{5}j^{4}|_{\gamma \rightarrow 0}
\end{eqnarray*}

Consider now the limit which introduced by Bianchi, Magliaro and Perini \cite{Bianchi:2009ri}, i.e. $
\gamma \rightarrow 0,\,j\rightarrow \infty $, with fixed physical area $\gamma j=A$. Then the only term that survives are Eq.\eqref{resta} and  Eq.\eqref{resta2}.  These terms are precisely those appearing in the Regge calculus three-point  function, given in \cite{Bianchi:RC2008}.

Therefore, we can conclude that in the Bianchi-Magliaro-Perini limit the 3 point function of loop quantum gravity matches the Regge calculus one.

With an analogous ``dimensional'' analysis, we can check that for 4-point function and 5-point function the spin foam model give perturbative Regge calculus result in the same limit. For 4-point function, the Regge part has the scale of $\mathcal{O}(\gamma^5j^5)$, others have the scale of $\mathcal{O}(\gamma^kj^5)$, $k>5$. For 4-point function, it is the same. The scale of Regge part is $\mathcal{O}(\gamma^6j^6)$.

It appears therefore that $\gamma$ scales the amplitude of the ``un-gluing" fluctuation. It also measures the difference between area bivectors $A^{IJ}$ and group generators $J^{IJ}$. The $\gamma\rightarrow0$ limit corresponds to $J^{IJ}=A^{IJ}$\cite{Freidel:2007py}\cite{Conrady:PI2008}.

  \section{Three-point  function in perturbative quantum gravity}\label{sec:3p}
In this section we give for completeness the analytic expression of the three-point function in position space, at tree level, in the harmonic gauge. We will briefly review the main definitions and notations on perturbative quantum general relativity, based on \cite{Zee:QFT}\cite{Donoghue:GR1994}\cite{Modesto:PG2003}. We only show the result in this note. More details are in the Appendix.
  \subsection{Definitions}
 perturbative quantum gravity describes the quantum gravitational field as a tensor field in a flat background spacetime. This is a weak field expansion that does not address the problem of the full consistency of the theory, but it gives nevertheless a credible approximation in the very low energy regime.  Therefore a consistent full theory of quantum gravity should match the perturbative results in the low energy limit.

Here we focus on the Euclidean spacetime and we take background spacetime to be flat; i.e. the metric of the background is $\delta_{\mu\nu}$. The definition of gravitation field $h_{\mu\nu}(x)$ is
  \begin{equation}\label{eq:h}
    h_{\mu\nu}(x)=g_{\mu\nu}(x)-\delta_{\mu\nu}
  \end{equation}
where $g_{\mu\nu}(x)$ is the total metric, $x$ is a cartesian coordinate which covers the background spacetime manifold.

Since we use a path integral formalism to write the quantum theory of perturbative gravitation field, we need rewrite Einstein-Hilbert (EH) action (without cosmological constant)
  \begin{equation}
    S=\frac{1}{16\pi G}\int\dd x\sqrt{g}R
  \end{equation}
in terms of the field $h_{\mu\nu}(x)$.   Under general coordinate transform the gravitation field $h_{\mu\nu}$ has a gauge freedom, with a structure similar to the electromagnetic field case. To compute the symmetric three-point function, we choose the harmonic gauge
  \begin{equation}\label{eq:TG}
    \p_{\mu}h^{\mu\nu}=\frac{1}{2}\p^{\nu}h
  \end{equation}
  where $h\equiv h_{\mu}^{\mu}$. We only consider the pure gravity situation, without matter. In this case, the linearization of the Einstein  equations reads
  \begin{equation}
    \partial _{\rho }\partial ^{\rho }h_{\mu \nu }=\frac{1}{2}\delta _{\mu \nu}\partial _{\rho }\partial ^{\rho }h.
  \end{equation}
Taking  the trace for both side, we have
  \begin{equation}\label{eq:EEh}
    \partial _{\rho }\partial ^{\rho }h=0,\quad \text{and} \quad \partial _{\rho }\partial ^{\rho }h_{\mu \nu }=0
  \end{equation}
Using this and the gauge fixing, the EH action becomes (only keeping the 3-valent terms)
  \begin{equation}\label{S3EH}
    S_3=\frac{1}{64\pi G}\int\dd x(h^{\sigma \rho}\partial_{\sigma}h^{\mu\nu}\partial_{\rho}h_{\mu\nu}-2h_{\mu\beta}\partial^{\sigma}h^{\mu
    \nu }\partial ^{\beta }h_{\nu\sigma })
  \end{equation}

  \subsection{Three-point  function}\label{sec:p3p}
The three-point function at the tree level leading order is defined as follow
  \begin{eqnarray}
    &&G_{\mu _{1}\mu _{2}\nu _{1}\nu _{2}\sigma _{1}\sigma_{2}}(x_{1},x_{2},x_{3}) =\\ \nonumber
    &&\frac{1}{Z}\int \D h~\ex^{\im S_{2}} ~\im S_3~h_{\mu _{1}\mu_{2}}\left( x_{1}\right) h_{\nu _{1}\nu _{2}}\left( x_{2}\right) h_{\sigma_{1}\sigma _{2}}\left( x_{3}\right)
  \end{eqnarray}
  where $Z =\int \D h~\exp \left( iS_{2}\right)$ and
  \begin{equation}
  S_{2}=\frac{1}{64\pi G}\int \dd^{4}z(\partial^{\sigma}h^{\mu\beta}\partial _{\sigma}h_{\beta \mu }-\frac{1}{2}\partial _{\rho }h\partial ^{\rho }h).
  \end{equation}
 The terms in $S_3$ are quite analogous; let's focus on the first, namely $h^{\sigma\rho}\partial_{\sigma}h^{\mu\nu}\partial_{\rho}h_{\mu\nu}$. Using the Wick contraction method, we obtain  \begin{eqnarray}\label{3Pp}
\nonumber&&  \!\!  G_{\mu _{1}\mu _{2}\nu _{1}\nu _{2}\sigma _{1}\sigma
_{2}}(x_{1},x_{2},x_{3}) =
\frac{i}{64\pi G}\frac{1}{Z}\int \D h\exp ( iS_{2})  \\
\nonumber&& \int\! d^{4}z\, h^{\sigma \rho}(z) \partial _{\sigma
}h^{\mu \nu }(z) \partial _{\rho }h_{\mu \nu }( z)
h_{\mu _{1}\mu _{2}}( x_{1}) h_{\nu _{1}\nu _{2}}(
x_{2}) h_{\sigma _{1}\sigma _{2}}( x_{3})  \\
&&=\frac{\kappa }{2}\int d^{4}z
 \\
\nonumber&&\quad
(D_{\text{ \ },\mu _{1}\mu _{2}}^{\sigma \rho
}(z\!-\!x_{1})\partial _{\sigma }D_{\text{ \ },\nu _{1}\nu _{2}}^{\mu \nu
}( z\!-\!x_{2}) \partial _{\rho }D_{\mu \nu ,\sigma _{1}\sigma
_{2}}( z-x_{3})  \\
\nonumber&&\quad+D_{\text{ \ },\mu _{1}\mu _{2}}^{\sigma \rho }(z\!-\!x_{1})\partial _{\sigma
}D_{\ \ ,\sigma _{1}\sigma _{2}}^{\mu \nu }( z\!-\!x_{3}) \partial
_{\rho }D_{\mu \nu ,\nu _{1}\nu _{2}}( z\!-\!x_{2})  \\
\nonumber&&\quad+D_{\text{ \ },\nu _{1}\nu _{2}}^{\sigma \rho }(z\!-\!x_{2})\partial _{\sigma
}D_{\ \ ,\sigma _{1}\sigma _{2}}^{\mu \nu }( z\!-\!x_{3}) \partial
_{\rho }D_{\mu \nu ,\mu _{1}\mu _{2}}( z\!-\!x_{1})  \\
\nonumber&&\quad+D_{\text{ \ },\nu _{1}\nu _{2}}^{\sigma \rho }(z\!-\!x_{2})\partial _{\sigma
}D_{\ \ ,\mu _{1}\mu _{2}}^{\mu \nu }( z\!-\!x_{1}) \partial _{\rho
}D_{\mu \nu ,\sigma _{1}\sigma _{2}}( z\!-\!x_{3})  \\
\nonumber&&\quad+D_{\text{ \ },\sigma _{1}\sigma _{2}}^{\sigma \rho }(z\!-\!x_{3})\partial
_{\sigma }D_{\ \ ,\mu _{1}\mu _{2}}^{\mu \nu }( z\!-\!x_{1}) \partial
_{\rho }D_{\mu \nu ,\nu _{1}\nu _{2}}( z\!-\!x_{2})  \\
\nonumber&&\ \ +D_{\text{ \ },\sigma _{1}\sigma _{2}}^{\sigma \rho }(z\!-\!x_{3})\partial
_{\sigma }D_{\ \ ,\nu _{1}\nu _{2}}^{\mu \nu }( z\!-\!x_{2}) \partial
_{\rho }D_{\mu \nu ,\mu _{1}\mu _{2}}( z\!-\!x_{1}))
 \end{eqnarray}
where $\kappa=\sqrt{32\pi G}$ and $D_{\mu\nu,\rho\sigma}(x-y)$ is graviton propagator in position space, which is
  \begin{eqnarray}\nonumber
     D_{\mu \nu ,\rho \sigma }\left( x-y\right)&=&\frac{1}{8\pi ^{2}}\frac{1}{\left\vert x-y\right\vert ^{2}}(\delta _{\mu
\rho }\delta _{\nu \sigma }+\delta _{\mu \sigma }\delta _{\nu \rho }-\delta
_{\mu \nu }\delta _{\rho \sigma })\\
&\equiv& \frac{1}{8\pi ^{2}}\frac{1}{\left\vert x-y\right\vert ^{2}}\Delta _{\mu \nu ,\rho \sigma }
  \end{eqnarray}
We do not write the non-connected terms because they equal to zero by gauge symmetry.

Since all the terms in Eq.(\ref{3Pp}) have a similar form, we  focus on the first one. This reads
  \begin{eqnarray}\label{3PT}\nonumber
      &&\!\!\!\!\!\int \! d^{4}z\,D_{\text{ },\mu _{1}\mu _{2}}^{\sigma \rho }\!(\!z-x_{1}\!)\,\partial
_{\sigma }D_{\text{  },\nu _{1}\nu _{2}}^{\mu \nu }\!( \!z-x_{2}\!)\,
\partial _{\rho }D_{\mu \nu ,\sigma _{1}\sigma _{2}}\!( \!z-x_{3}\!)  \\
&&=\frac{1}{2\left( 2\pi \right) ^{6}}\int d^{4}z\frac{1}{\left\vert
z-x_{1}\right\vert ^{2}}\frac{z_{\sigma }-\left( x_{2}\right) _{\sigma }}{%
\left\vert z-x_{2}\right\vert ^{4}}\frac{z_{\rho }-\left( x_{3}\right)
_{\rho }}{\left\vert z-x_{3}\right\vert ^{4}}
\\ && \quad\quad\quad\quad\quad\quad
\Delta _{\text{ \ },\mu _{1}\mu
_{2}}^{\sigma \rho }\Delta _{\text{ \ },\nu _{1}\nu _{2}}^{\mu \nu }\Delta
_{\mu \nu ,\sigma _{1}\sigma _{2}}\nonumber
  \end{eqnarray}
  The difficulty is to solve the integral in Eq.(\ref{3PT}). The asymmetric form of the integral comes from the derivatives in the perturbative EH action (\ref{S3EH}). Fortunately we can change the derivative variables and take the derivatives out of the integral, turning it into a three-point  function in $\lambda\phi^3$ theory. For Eq.(\ref{3PT}), it turns into
  \begin{equation}
\frac{\Delta _{\text{ \ },\mu _{1}\mu _{2}}^{\sigma \rho }\Delta _{\text{
\ },\nu _{1}\nu _{2}}^{\mu \nu }\Delta _{\mu \nu ,\sigma _{1}\sigma _{2}}}{%
2\left( 2\pi \right) ^{6}}\frac{\partial }{\partial x_{2}^{\sigma }}\frac{%
\partial }{\partial x_{3}^{\rho }}G_{\lambda\phi^3}\left(x_1,x_2,x_3\right)
  \end{equation}
where
  \begin{equation}
G_{\lambda\phi^3}\left(x_1,x_2,x_3\right)=\int \frac{d^{4}z}{\left\vert
z-x_{1}\right\vert ^{2}\left\vert z-x_{2}\right\vert ^{2}\left\vert
z-x_{3}\right\vert ^{2}}\\
  \end{equation}
According to a theorem in \cite{Francesco:CFT}, for a scalar three-point function, which is rotation, translation  and dilation covariant, must have the form $G\left( x_{1},x_{2},x_{3}\right) =Cx_{12}^{\alpha }x_{23}^{\beta}x_{31}^{\gamma }$ in general, where $x_{ij}=\left\vert x_{i}-x_{j}\right\vert $, $C$ is a constant. Then
  \begin{equation}
    G_{\lambda\phi^3}\left(x_1,x_2,x_3\right)=\frac{C}{\left\vert x_{1}-x_{2}\right\vert ^{\frac{2}{3}}\left\vert
x_{2}-x_{3}\right\vert ^{\frac{2}{3}}\left\vert x_{3}-x_{1}\right\vert ^{%
\frac{2}{3}}}
  \end{equation}
Then the derivatives outside of the integral give the final results. Let us introduce some notations. Focus on an equilateral
4-simplex. $\left\vert x_{1}-x_{2}\right\vert =\left\vert
x_{2}-x_{3}\right\vert =\left\vert x_{3}-x_{1}\right\vert =L$, $%
x_{1}^{2}=x_{2}^{2}=x_{3}^{2}=\frac{2}{5}L^{2}$ and $x_{i}\cdot
x_{j}|_{i\neq j}=-\frac{1}{10}L^{2}$. Writing
\begin{equation}
I_{ij}^{\mu\nu }=\frac{\partial}{\partial _{\mu }x_{i}}\frac{\partial}{\partial _{\nu }x_{j}} G_{\lambda\phi^3}\left(x_1,x_2,x_3\right),
\end{equation} we have, for instance
\begin{widetext}
  \begin{equation}\label{eq:Is}
I_{12}^{\mu \nu } =C\frac{4}{9L^{6}}(x_{3}^{\mu }x_{3}^{\nu }+x_{3}^{\mu
}x_{1}^{\nu }-2x_{3}^{\mu }x_{2}^{\nu }-2x_{1}^{\mu }x_{3}^{\nu
}-5x_{1}^{\mu }x_{1}^{\nu }+7x_{1}^{\mu }x_{2}^{\nu }+x_{2}^{\mu }x_{3}^{\nu
}+4x_{2}^{\mu }x_{1}^{\nu }-5x_{2}^{\mu }x_{2}^{\nu })
\end{equation}
and similarly for the other components. This allows us to write the three-point  function explicitly:
  \begin{equation}\label{eq:3pp}
  \begin{split}
    G_{\mu _{1}\mu _{2}\nu _{1}\nu _{2}\sigma _{1}\sigma
_{2}}(x_{1},x_{2},x_{3})&=\frac{\kappa }{2}\frac{1}{2\left( 2\pi \right) ^{6}}\bigg(\left(
\begin{array}{c}
I_{23}^{\sigma \rho }\Delta _{\sigma \rho ,\mu _{1}\mu _{2}}\Delta _{\mu \nu
,\nu _{1}\nu _{2}}\Delta _{\mu \nu ,\sigma _{1}\sigma _{2}}+I_{32}^{\sigma
\rho }\Delta _{\sigma \rho ,\mu _{1}\mu _{2}}\Delta _{\mu \nu ,\sigma
_{1}\sigma _{2}}\Delta _{\mu \nu ,\nu _{1}\nu _{2}} \\
+I_{31}^{\sigma \rho }\Delta _{\sigma \rho ,\nu _{1}\nu _{2}}\Delta _{\mu
\nu ,\sigma _{1}\sigma _{2}}\Delta _{\mu \nu ,\mu _{1}\mu
_{2}}+I_{13}^{\sigma \rho }\Delta _{\sigma \rho ,\nu _{1}\nu _{2}}\Delta
_{\mu \nu ,\mu _{1}\mu _{2}}\Delta _{\mu \nu ,\sigma _{1}\sigma _{2}} \\
+I_{12}^{\sigma \rho }\Delta _{\sigma \rho ,\sigma _{1}\sigma _{2}}\Delta
_{\mu \nu ,\mu _{1}\mu _{2}}\Delta _{\mu \nu ,\nu _{1}\nu
_{2}}+I_{21}^{\sigma \rho }\Delta _{\sigma \rho ,\sigma _{1}\sigma
_{2}}\Delta _{\mu \nu ,\nu _{1}\nu _{2}}\Delta _{\mu \nu ,\mu _{1}\mu _{2}}%
\end{array}%
\right)\\
&\quad-2\left(
\begin{array}{c}
I_{23}^{\sigma \beta }\Delta _{\mu \beta ,\mu _{1}\mu _{2}}\Delta _{\mu \nu
,\nu _{1}\nu _{2}}\Delta _{\nu \sigma ,\sigma _{1}\sigma
_{2}}+I_{32}^{\sigma \beta }\Delta _{\mu \beta ,\mu _{1}\mu _{2}}\Delta
_{\mu \nu ,\sigma _{1}\sigma _{2}}\Delta _{\nu \sigma ,\nu _{1}\nu _{2}} \\
+I_{31}^{\sigma \beta }\Delta _{\mu \beta ,\nu _{1}\nu _{2}}\Delta _{\mu \nu
,\sigma _{1}\sigma _{2}}\Delta _{\nu \sigma ,\mu _{1}\mu
_{2}}+I_{13}^{\sigma \beta }\Delta _{\mu \beta ,\nu _{1}\nu _{2}}\Delta
_{\mu \nu ,\mu _{1}\mu _{2}}\Delta _{\nu \sigma ,\sigma _{1}\sigma _{2}} \\
+I_{12}^{\sigma \beta }\Delta _{\mu \beta ,\sigma _{1}\sigma _{2}}\Delta
_{\mu \nu ,\mu _{1}\mu _{2}}\Delta _{\nu \sigma ,\nu _{1}\nu
_{2}}+I_{21}^{\sigma \beta }\Delta _{\mu \beta ,\sigma _{1}\sigma
_{2}}\Delta _{\mu \nu ,\nu _{1}\nu _{2}}\Delta _{\nu \sigma ,\mu _{1}\mu
_{2}}%
\end{array}%
\right)\bigg)
  \end{split}
  \end{equation}
\end{widetext}
  \subsection{Comparison between the perturbative and loop three-point  functions}\label{sec:SFvsP}

The comparison of the three-point function computed here with the one computed in the
previous section is not easy.  In order to compare the expectation values, we need to identify the Penrose operators $G_l^{ab}$ with the metric field. The Penrose operator has a clear geometrical interpretation \cite{Rovelli:2010km}: it is the scalar product of the flux operator across the boundary triangles $a$ and $b$ of the boundary tetrahedron $l$ of a 4-simplex-like spacetime region. It can therefore immediately compared with quantities well defined in Regge geometry: areas of triangles and angles between triangles.

The direct comparison with the metric operator, on the other hand, is tricky, since areas and angles of simplices are nonlocal functions of the metric. In addition, the $n$-point functions are computed in the linearized theory in a certain gauge. The loop theory defines implicitly a gauge in two steps. First, the boundary operators are naturally defined in a ``time" gauge, with respect to the foliation defined by the boundary. Second, the remaining gauge freedom is fixed by the boundary state \cite{Speziale:2008uw,Dittrich:2007wm}.

Tentatively, we may write
      \begin{equation}
        G_n^{ab}=E_n^a\cdot E_n^b=\det(q)q^{ij}(x)N^{na}_i(x)N^{nb}_j(x)
      \end{equation}
where $N^{na}_i$ is the normal one form to the triangle $(n,a)$ in the plane of the tetrahedron $a$, normalized to the coordinate area of the triangle, in the background geometry, and $q_{ij}$ is the three metric induced on the boundary.
More precisely, we can use the two-form $B^{la}_{\mu\nu}$ associated to the $(n,a)$ triangle and write
   \begin{equation}
G_n^{ab}=2g^{\rho\sigma}g^{\mu\nu}B^{la}_{\rho\mu}B^{lb}_{\sigma\nu}.
   \end{equation}
 This is the way the loop operator was identified with the perturbative gravitational field in \cite{Bianchi:2009ri}. The same simple minded identification does not appear to work for the three-point  function, as shown by an explicit numerical calculation given in the Appendix \ref{numerical}, if we use the numerical values for the boundary state found in  \cite{Bianchi:2009ri}.
Since the loop calculation matches the Regge one, the inconsistency is not related to the specific of the loop formalism, and is therefore of secondary interest here.

The problem of the consistency between Regge calculus \cite{Regge:1961px} and continuum perturbative quantum gravity field theory has been discussed in  \cite{Friedberg:1984ma,Feinberg:1984he,Hamber:QG-Path}.   The consistency between Regge calculus and continuum theory is based on the relation between the Regge action $S_{\text{Regge}}$ and EH action $S_{\text{EH}}$. $S_{\text{Regge}}$ can be derived from $S_{\text{EH}}$ \cite{Friedberg:1984ma}, and $S_{\text{Regge}}$ yields back  $S_{\text{EH}}$ with a correction in the order $\mathcal{O}(l^2/\rho^2)$ \cite{Feinberg:1984he}, where $l$ is the typical length of a four simplex and $\rho$ is Gauss radius which stands for the intrinsic curvature. In the limit $l\rightarrow0$ or $\rho\rightarrow\infty$, $S_{\text{Regge}}\rightarrow S_{\text{EH}}$. In our calculation, we use the limit $\rho\rightarrow\infty$, as we have mentioned in Section \ref{sec:VA}. Then we can use the regular way to calculate the graviton $n$-point function, i.e. adding $n$ $h_{\mu\nu}$s into the path integral as insertions and change the action $S_{\text{EH}}\rightarrow S_{\text{EH}}+\mathcal{O}(l^2/\rho^2)$\cite{Feinberg:1984he}.  Perturbative Regge calculus is given by the strong coupling expansion \cite{Hamber:QG-Path}.  The expansion around the saddle point in loop gravity corresponds to the strong coupling expansion in Regge calculus.

We also point out here that in \cite{Rovelli:2005yj}\cite{Bianchi:2006uf}, the traceless gauge $h_{\mu}^{\mu}=0$ was assumed, but this may not be consistent with the gauge choice implicit in the use of the Penrose field operator. If we take this into account in the definition of two-point  function given in \cite{Bianchi:2009ri}
  \begin{equation}
    G_{mn}^{abcd}=\langle E_m^a\cdot E_m^b~E_n^c\cdot E_n^d\rangle-\langle E_m^a\cdot E_m^b\rangle\langle E_n^c\cdot E_n^d\rangle
  \end{equation}
since $E_n^a$ is a densitized operator, we obtain
  \begin{eqnarray}\nonumber
    G_{mn}^{abcd}&=\langle \det(g(x_m))g_{\mu\nu}(x_m)~\det(g(x_n))g_{\rho\sigma}(x_n)\rangle \\ \nonumber&\quad\quad(N_m^a)^{\mu}(N_m^b)^{\nu}(N_n^c)^{\rho}(N_n^d)^{\sigma}\\ \nonumber
    &\quad-\langle \det(g(x_m))g_{\mu\nu}(x_m)\rangle\langle\det(g(x_n))g_{\rho\sigma}(x_n)\rangle \\&\quad\quad (N_m^a)^{\mu}(N_m^b)^{\nu}(N_n^c)^{\rho}(N_n^d)^{\sigma}
  \end{eqnarray}
  Using (\ref{eq:gtoh}), we find at the order $\mathcal{O}(h^2)$
  \begin{eqnarray}\nonumber
&    G_{mn}^{abcd}=\langle hh_{\rho \sigma }\delta _{\alpha \beta }+h^{2}\delta
_{\rho \sigma }\delta _{\alpha \beta }+hh_{\alpha \beta }\delta _{\rho
\sigma }+h_{\rho \sigma }h_{\alpha \beta }\rangle
\\&\quad\quad  (N_m^a)^{\mu}(N_m^b)^{\nu}(N_n^c)^{\rho}(N_n^d)^{\sigma} +\mathcal{O}(h^3)
  \end{eqnarray}
 which is certainly not the standard two-point  function. For the three-point  function case, the relation is even more complicated.

An additional source of uncertainty in the relation between the flux variables $E_n^a$  and $g_{\mu\nu}$ is given by the correct identification of the normals. Above we have assumed
      \begin{equation}
        E_n^aE_n^b=\det(g)g_{\mu\nu}(x)N_n^a(x)N_n^b(x)
      \end{equation}
      where the normals $N_n^a$  are those of the background geometry. But in the boundary state used $N_n^a=j_{na}n_n^a(j(h))$, where the normals are determined by the areas of the entire 4-simplex.  This gives an extra dependence on the metric: $\det(g)g_{\mu\nu}(x)N_n^a(j(h(x)))N_n^b(j(h(x)))$.

Because of these various technical complications a direct comparison with the weak field expansion in $g_{\mu\nu}$ requires more work.  On the other hand, it is not clear that this work is of real interest, since the key result of the consistency of the loop dynamics with the Regge one is already established.

  \section{Conclusion}

We have computed the three-point  function of loop quantum gravity, starting from the background independent spinfoam dynamics, at the lowest order in the vertex expansion. We have shown that this is equivalent to the one of perturbative Regge calculus in the limit $\gamma\rightarrow0$, $j\rightarrow\infty$ and $\gamma j=A$.

Given the good indications on the large distance limit of the $n$-point functions for Euclidean quantum gravity, we think the most urgent open problem is to extend these results to the Lorentzian case, and to the theory with matter \cite{Bianchi:2010bn,Han:2011as} and cosmological constant \cite{Fairbairn:2010cp,Han:2010pz,Bianchi:2011ym}.

Among the problem that we leave open, are the following. (i) We have computed the  three-point  function in position space from perturbative quantum gravity, treated as a flat-space quantum field theory. We have found that we cannot use here the techniques of \cite{Rovelli:2005yj,Bianchi:2006uf,Alesci:2007tx,Bianchi:2009ri} to compare this with the loop calculation, because of technical complications in comparing the expansion.  These can be traced to the different gauges in which the calculations are performed, to the traceless condition $h_{\mu}^{\mu}=0$ which in general is not satisfied and to the fact that the normals have a non-trivial relation with the field $N_n^a=N_n^a(h)$.
(ii) The boundary vacuum state and the parameters $\alpha^{(ab)(cd)}$ introduced in Eq.(\ref{eq:VSc}) should be better understood and checked. A possibility is to compute them from the first principle, using the unitary condition $\braket{W}{\Psi_{\gamma}}=1$.\footnote{private communication with Simone Speziale}.
 (iii) The gauge implicit in the use of the loop formalism is not completely clear.
    In weak field expansion, the De Donder-like  (harmonic) gauge, turns out to be consistent for the lattice graviton propagator \cite{Rocek:1982fr,Hamber:QG-Path}, and with the radial structure of the loop calculation \cite{Magliaro:2007qr}. But the extension of this to higher $n$-point functions in not clear.

  \section{Acknowledgments}
Thanks to Eugenio Bianchi, Muxin Han, Simone Speziale, Wolfgang Wieland for continuous helps, discussions and advices.  Additional thanks to Eugenio Bianchi for sharing his notes and files. M.Z. is supported by CSC scholarship No.2010601003.

  \appendix
  \section{From EH action to Eq.(\ref{S3EH})}
We follow the work of Modesto \cite{Modesto:PG2003} and Bianchi and Modesto \cite{Bianchi:2007vf}.   We split the total metric $g_{\mu\nu}$ into the background metric $\delta_{\mu\nu}$ and the fluctuation $h_{\mu\nu}$, as in (\ref{eq:h}). Form this we can get its inverse $g^{\mu\nu}$ and square root of its determinant $\sqrt{\left\vert \det \left( g\right) \right\vert }$
  \begin{eqnarray}\label{eq:gtoh}
    g^{\mu\nu}&=&\sum_{n=0}^{\infty }\left(
-\right) ^{n}\left( h^{n}\right) ^{\mu \nu }\\
    \sqrt{\left\vert \det \left( g\right) \right\vert }&=&\prod_{k=1}^{\infty }\left( \sum_{m=0}^{\infty }\frac{%
\left( -\right) ^{m\left( k+1\right) }}{m!\left( 2k\right) ^{m}}\left(
\left( h^{k}\right) _{\kappa }^{\kappa }\right) ^{m}\right)
  \end{eqnarray}
  Because EH action contains Ricci scalar and Ricci scalar contains Christopher symbol, we write the symbol based on $h_{\mu\nu}$.
  \begin{equation}
    \Gamma _{\nu \mu }^{\sigma } =\frac{1}{2}\sum_{n_{1}=0}^{\infty }\left(
-\right) ^{n_{1}}\left( h^{n_{1}}\right) ^{\sigma \rho }\left( \partial
_{\mu }h_{\nu \rho }+\partial _{\nu }h_{\rho \mu }-\partial _{\rho }h_{\nu
\mu }\right) \\
  \end{equation}

\begin{widetext}
  Then we can finally get the Lagrangian based on $h_{\mu\nu}$
  \begin{equation}\label{eq:ELh}
    \begin{split}
      L &=\frac{1}{16\pi G}\sqrt{\left\vert \det \left( g\right) \right\vert }R \\
&=\frac{1}{16\pi G}\prod_{k=1}^{\infty }\left( \sum_{m=0}^{\infty }\frac{%
\left( -\right) ^{m\left( k+1\right) }}{m!\left( 2k\right) ^{m}}\left(
\left( h^{k}\right) _{\kappa }^{\kappa }\right) ^{m}\right) \\
&\quad\times (\frac{1}{2}\sum_{n_{1}=0}^{\infty }\sum_{n=0}^{\infty }\left(
-\right) ^{n+n_{1}}\left( h^{n}\right) ^{\mu \nu }(\partial _{\sigma }\left(
h^{n_{1}}\right) ^{\sigma \rho }\partial _{\mu }h_{\nu \rho }+\partial
_{\sigma }\left( h^{n_{1}}\right) ^{\sigma \rho }\partial _{\nu }h_{\rho \mu
}-\partial _{\sigma }\left( h^{n_{1}}\right) ^{\sigma \rho }\partial _{\rho
}h_{\nu \mu } \\
&\quad+\left( h^{n_{1}}\right) ^{\sigma \rho }\partial _{\sigma }\partial _{\mu
}h_{\nu \rho }+\left( h^{n_{1}}\right) ^{\sigma \rho }\partial _{\sigma
}\partial _{\nu }h_{\rho \mu }-\left( h^{n_{1}}\right) ^{\sigma \rho
}\partial _{\sigma }\partial _{\rho }h_{\nu \mu }) \\
&\quad-\frac{1}{2}\sum_{n_{2}=0}^{\infty }\sum_{n=0}^{\infty }\left( -\right)
^{n+n_{2}}\left( h^{n}\right) ^{\mu \nu }(\partial _{\nu }\left(
h^{n_{2}}\right) ^{\sigma \rho }\partial _{\mu }h_{\sigma \rho }+\partial
_{\nu }\left( h^{n_{2}}\right) ^{\sigma \rho }\partial _{\sigma }h_{\rho \mu
}-\partial _{\nu }\left( h^{n_{2}}\right) ^{\sigma \rho }\partial _{\rho
}h_{\sigma \mu } \\
&\quad+\left( h^{n_{2}}\right) ^{\sigma \rho }\partial _{\nu }\partial _{\mu
}h_{\sigma \rho }+\left( h^{n_{2}}\right) ^{\sigma \rho }\partial _{\nu
}\partial _{\sigma }h_{\rho \mu }-\left( h^{n_{2}}\right) ^{\sigma \rho
}\partial _{\nu }\partial _{\rho }h_{\sigma \mu }) \\
&\quad+\frac{1}{4}\sum_{n_{3}=0}^{\infty }\sum_{n_{4}=0}^{\infty
}\sum_{n=0}^{\infty }\left( -\right) ^{n+n_{3}+n_{4}}\left( h^{n}\right)
^{\mu \nu }\left( h^{n_{3}}\right) ^{\sigma \alpha }\left( h^{n_{4}}\right)
^{\rho \beta }\left( \partial _{\rho }h_{\sigma \alpha }+\partial _{\sigma
}h_{\alpha \rho }-\partial _{\alpha }h_{\sigma \rho }\right) \\
&\quad\times \left( \partial _{\mu }h_{\nu \beta }+\partial _{\nu }h_{\beta \mu
}-\partial _{\beta }h_{\nu \mu }\right) \\
&\quad-\frac{1}{4}\sum_{n_{5}=0}^{\infty }\sum_{n_{6}=0}^{\infty
}\sum_{n=0}^{\infty }\left( -\right) ^{n+n_{5}+n_{6}}\left( h^{n}\right)
^{\mu \nu }\left( h^{n_{5}}\right) ^{\sigma \alpha }\left( h^{n_{6}}\right)
^{\rho \beta }\left( \partial _{\rho }h_{\nu \alpha }+\partial _{\nu
}h_{\alpha \rho }-\partial _{\alpha }h_{\nu \rho }\right) \\
&\quad\times \left( \partial _{\mu }h_{\sigma \beta }+\partial _{\sigma
}h_{\beta \mu }-\partial _{\beta }h_{\sigma \mu }\right) )
    \end{split}
  \end{equation}

Before simplify the Lagrangian to Eq.(\ref{S3EH}), we  emphasize that we use the harmonic gauge (\ref{eq:TG}) and the Einstein equation without matter field (\ref{eq:EEh}). We also assume that the integral over a total derivative vanishes. Under these assumptions, we find that the terms like $h_{\mu \nu }\partial ^{\rho }h^{\mu \nu }\partial _{\rho }h,h\partial ^{\rho}h\partial _{\rho }h,h_{\nu }^{\mu }\partial _{\rho }h_{\mu \sigma }\partial^{\rho }h^{\nu \sigma }$ vanish after integration in the action. We use $h_{\mu \nu }\partial ^{\rho }h^{\mu \nu }\partial _{\rho }h$ as and example to prove this.
  \begin{eqnarray*}
\int d^{4}xh_{\mu \nu }\partial ^{\rho }h^{\mu \nu }\partial _{\rho }h
&=&\int d^{4}x\frac{1}{2}\left( h_{\mu \nu }\partial ^{\rho }h^{\mu \nu
}\partial _{\rho }h+h^{\mu \nu }\partial ^{\rho }h_{\mu \nu }\partial _{\rho
}h\right)  \\
&=&\int d^{4}x\frac{1}{2}\partial ^{\rho }\left( h_{\mu \nu }h^{\mu \nu
}\right) \partial _{\rho }h \\
&=&-\int d^{4}x\frac{1}{2}h_{\mu \nu }h^{\mu \nu }\partial ^{\rho }\partial
_{\rho }h \\
&=&0
\end{eqnarray*}

  Now we can simplify Eq.(\ref{eq:ELh}). For the three-point  function, we just need the terms with the form $h\partial h\partial h$. We denote it $L_{3}$ and simplify it case by case.
  For $km=0$
  \begin{equation*}
    \prod_{k=1}^{\infty }\left( \sum_{m=0}^{\infty }\frac{\left( -\right)
^{m\left( k+1\right) }}{m!\left( 2k\right) ^{m}}\left( \left( h^{k}\right)
_{\kappa }^{\kappa }\right) ^{m}\right) =\prod_{k=1}^{\infty }\left( \frac{%
\left( -\right) ^{0\left( k+1\right) }}{0!\left( 2k\right) ^{0}}\left(
\left( h^{k}\right) _{\kappa }^{\kappa }\right) ^{0}\right)
=\prod_{k=1}^{\infty }1=1
  \end{equation*}
  Then
  \begin{equation*}
      L_{3}^{\left( 0\right) } =\frac{1}{16\pi G} =\frac{1}{64\pi G}(h^{\sigma \rho }\partial _{\sigma }h^{\mu \nu }\partial
_{\rho }h_{\mu \nu }-2h^{\mu \nu }\partial _{\nu }h_{\beta \mu }\partial
^{\beta }h+2h^{\mu \nu }\partial ^{\beta }h_{\nu \sigma }\partial ^{\sigma
}h_{\mu \beta })
  \end{equation*}

  For $km=1$
  \begin{equation*}
    \prod_{k=1}^{\infty }\left( \sum_{m=0}^{\infty }\frac{\left( -\right)
^{m\left( k+1\right) }}{m!\left( 2k\right) ^{m}}\left( \left( h^{k}\right)
_{\kappa }^{\kappa }\right) ^{m}\right) =\frac{\left( -\right) ^{1\left(
1+1\right) }}{1!\left( 2\right) ^{1}}h_{\kappa }^{\kappa }=\frac{1}{2}h
  \end{equation*}
  Then
  \begin{eqnarray*}
L_{3}^{\left( 1\right) } &=&\frac{1}{16\pi G}\frac{1}{2}h =-\frac{1}{64\pi G}h\partial _{\rho }h_{\nu \alpha }\partial ^{\nu
}h^{\alpha \rho }
\end{eqnarray*}

  For $km=2$
  \begin{eqnarray*}
\prod_{k=1}^{\infty }\left( \sum_{m=0}^{\infty }\frac{\left( -\right)
^{m\left( k+1\right) }}{m!\left( 2k\right) ^{m}}\left( \left( h^{k}\right)
_{\kappa }^{\kappa }\right) ^{m}\right) &=&\frac{\left( -\right) ^{2\left(
1+1\right) }}{2!\left( 2\right) ^{2}}\left( h_{\kappa }^{\kappa }\right)
^{2}+\frac{\left( -\right) ^{1\left( 2+1\right) }}{1!\left( 4\right) ^{1}}%
\left( h^{2}\right) _{\kappa }^{\kappa } \\
&=&\frac{1}{8}h^{2}-\frac{1}{4}h^{\kappa \lambda }h_{\kappa \lambda }
\end{eqnarray*}
  Then
  \begin{eqnarray*}
L_{3}^{\left( 2\right) } =0
\end{eqnarray*}

  Thus we can get $L_{3}$
  \begin{equation}
    \begin{split}
      L_{3} &=\frac{1}{64\pi G}(h^{\sigma \rho }\partial _{\sigma }h^{\mu \nu
}\partial _{\rho }h_{\mu \nu }-2h^{\mu \nu }\partial _{\nu }h_{\beta \mu
}\partial ^{\beta }h+2h^{\mu \nu }\partial ^{\beta }h_{\nu \sigma }\partial
^{\sigma }h_{\mu \beta }-h\partial _{\rho }h_{\nu \alpha }\partial ^{\nu
}h^{\alpha \rho }) \\
&=\frac{1}{64\pi G}(h^{\sigma \rho }\partial _{\sigma }h^{\mu \nu }\partial
_{\rho }h_{\mu \nu }+h\partial ^{\beta }h^{\mu \nu }\partial _{\nu }h_{\beta
\mu }-h^{\mu \nu }\partial _{\nu }h\partial _{\mu }h+2h^{\mu \nu }\partial
^{\beta }h_{\nu \sigma }\partial ^{\sigma }h_{\mu \beta }) \\
&=\frac{1}{64\pi G}\left( -h^{\nu \mu }\partial _{\mu }h_{\nu \beta
}\partial ^{\beta }h-\frac{1}{2}h^{\mu \nu }\partial _{\mu }h\partial _{\nu
}h+h^{\sigma \rho }\partial _{\sigma }h^{\mu \nu }\partial _{\rho }h_{\mu
\nu }+2h^{\mu \nu }\partial ^{\sigma }h_{\mu \beta }\partial ^{\beta }h_{\nu
\sigma }\right)  \\
&=\frac{1}{64\pi G}\left( h^{\sigma \rho }\partial _{\sigma }h^{\mu \nu
}\partial _{\rho }h_{\mu \nu }-2h_{\mu \beta }\partial ^{\sigma }h^{\mu \nu
}\partial ^{\beta }h_{\nu \sigma }\right)
    \end{split}
  \end{equation}
  This is  (\ref{S3EH}).

  \section{The total analytical expression of the derivatives in Sec.\ref{sec:AE}}
     The second order of the total action
    \begin{eqnarray}
      \frac{\p^2 S_{\text{tot}}}{\p j_{ab}\p j_{cd}}\bigg|_{\theta=0}&=&-\frac{\gamma\alpha^{(ab)(cd)}}{\sqrt{j_{0ab}}\sqrt{j_{0cd}}}+\im\frac{\p^2 S_{\text{Regge}}}{\p j_{ab}\p j_{cd}}\\
      \frac{\partial ^{2}S}{\partial \theta _{j}^{a\pm }\partial \theta _{i}^{a\pm
}}\bigg|_{\theta =0}&=&-\frac{1}{2}\gamma ^{\pm }\sum_{\left( b\neq a\right)
}j_{ab}\left( \delta _{ij}-\left( n_{ab}^{\pm }\right) _{i}\left(
n_{ab}^{\pm }\right) _{j}\right)\\
\frac{\partial ^{2}S}{\partial \theta _{j}^{b\pm }\partial \theta _{i}^{a\pm
}}\bigg|_{\theta =0}&=&\frac{1}{2}\gamma ^{\pm }j_{ab}(\delta _{ij}-\left(
n_{ab}^{\pm }\right) _{i}\left( n_{ab}^{\pm }\right) _{j}-i\varepsilon
_{ijk}\left( n_{ab}^{\pm }\right) _{k})
    \end{eqnarray}
     The third order of the total action
    \begin{eqnarray}
      \frac{\p^3 S_{\text{tot}}}{\p j_{ab}\p j_{cd}\p j_{ef}}\bigg|_{\theta=0}&=&\im\frac{\p^3 S_{\text{Regge}}}{\p j_{ab}\p j_{cd}\p j_{ef}}
         \end{eqnarray}
            \begin{eqnarray}
      \frac{\partial ^{3}S}{\partial \theta _{k}^{a\pm }\partial \theta _{j}^{a\pm
}\partial \theta _{i}^{a\pm }}\bigg|_{\theta =0}&=&\sum_{b\neq a}\frac{1}{6}i\gamma
^{\pm }j_{ab}(\delta _{jk}\left( n_{ab}^{\pm }\right) _{i}+\delta
_{ki}\left( n_{ab}^{\pm }\right) _{j}+\allowbreak \delta _{ij}\left(
n_{ab}^{\pm }\right) _{k}-3\left( n_{ab}^{\pm }\right) _{i}\left(
n_{ab}^{\pm }\right) _{j}\left( n_{ab}^{\pm }\right) _{k})
   \end{eqnarray}
            \begin{eqnarray}
\frac{\partial ^{3}S}{\partial \theta _{k}^{b\pm }\partial \theta _{j}^{a\pm
}\partial \theta _{i}^{a\pm }}|_{\theta =0} &=&-\frac{1}{4}i\gamma ^{\pm }j_{ab}(\delta _{jk}\left( n_{ab}^{\pm }\right)
_{i}+\delta _{ki}\left( n_{ab}^{\pm }\right) _{j}-2\left( n_{ab}^{\pm
}\right) _{i}\left( n_{ab}^{\pm }\right) _{j}\left( n_{ab}^{\pm }\right)
_{k}\\ \nonumber
&&+i\left( \varepsilon _{kil}\left( n_{ab}^{\pm }\right) _{j}+\varepsilon
_{kjl}\left( n_{ab}^{\pm }\right) _{i}\right) \left( n_{ab}^{\pm }\right)
_{l})   \end{eqnarray}
            \begin{eqnarray}
\frac{\partial ^{3}S}{\partial \theta _{k}^{a\pm }\partial \theta _{j}^{b\pm
}\partial \theta _{i}^{a\pm }}|_{\theta =0} &=&-\frac{1}{4}i\gamma ^{\pm }j_{ab}(\delta _{jk}\left( n_{ab}^{\pm }\right)
_{i}+\allowbreak \delta _{ij}\left( n_{ab}^{\pm }\right) _{k}-2\left(
n_{ab}^{\pm }\right) _{i}\left( n_{ab}^{\pm }\right) _{j}\left( n_{ab}^{\pm
}\right) _{k}\\ \nonumber
&&+i\varepsilon _{jil}\left( n_{ab}^{\pm }\right) _{l}\left(
n_{ab}^{\pm }\right) _{k}+i\varepsilon _{jkl}\left( n_{ab}^{\pm }\right)
_{l}\left( n_{ab}^{\pm }\right) _{i})
   \end{eqnarray}
            \begin{eqnarray}
\frac{\partial ^{3}S}{\partial \theta _{k}^{a\pm }\partial \theta _{j}^{a\pm
}\partial \theta _{i}^{b\pm }}|_{\theta =0} &=&-\frac{1}{4}i\gamma ^{\pm }j_{ab}(\delta _{ki}\left( n_{ab}^{\pm }\right)
_{j}+\delta _{ij}\left( n_{ab}^{\pm }\right) _{k}-2\left( n_{ab}^{\pm
}\right) _{i}\left( n_{ab}^{\pm }\right) _{j}\left( n_{ab}^{\pm }\right)
_{k}\\ \nonumber
&&+i\varepsilon _{ijl}\left( n_{ab}^{\pm }\right) _{l}\left( n_{ab}^{\pm
}\right) _{k}+i\varepsilon _{ikl}\left( n_{ab}^{\pm }\right) _{l}\left(
n_{ab}^{\pm }\right) _{j})
   \end{eqnarray}
            \begin{eqnarray}
\frac{\partial ^{e}S}{\p j_{cd}\partial \theta _{j}^{a\pm }\partial \theta _{i}^{a\pm
}}\bigg|_{\theta =0}&=&-\frac{1}{2}\gamma ^{\pm }\frac{\p}{\p j_{cd}}\left(\sum_{\left( b\neq a\right)
}j_{ab}\left( \delta _{ij}-\left( n_{ab}^{\pm }\right) _{i}\left(
n_{ab}^{\pm }\right) _{j}\right)\right)
   \end{eqnarray}
            \begin{eqnarray}
\frac{\partial ^{3}S}{\p j_{cd}\partial \theta _{j}^{b\pm }\partial \theta _{i}^{a\pm
}}\bigg|_{\theta =0}&=&\frac{1}{2}\gamma ^{\pm }\frac{\p}{\p j_{cd}}\left(j_{ab}(\delta _{ij}-\left(
n_{ab}^{\pm }\right) _{i}\left( n_{ab}^{\pm }\right) _{j}-i\varepsilon
_{ijk}\left( n_{ab}^{\pm }\right) _{k})\right)
    \end{eqnarray}
     The first derivatives of the insertions
        \begin{eqnarray}
          \frac{\partial q_{c}^{ab}}{\partial j_{ef}} &=&\gamma ^{2}\frac{\partial
(j_{ca}j_{cb}n_{ca}\cdot n_{cb})}{\partial j_{ef}}=\gamma ^{2}\frac{\partial
(j_{ca}j_{cb}\cos \theta _{cab})}{\partial j_{ef}} \\
\frac{\partial q_{n}^{ab}}{\partial \theta _{i}^{a\pm }}|_{\Theta _{i}^{a\pm
}=0}&=&-\frac{1}{2}i\gamma ^{2}\gamma ^{\pm }j_{na}j_{nb}(\left( n_{nb}^{\pm
}\right) _{i}-\left( n_{na}^{\pm }\right) _{i}\left( n_{nb}\right)
_{j}\left( n_{na}\right) _{j}+i\varepsilon _{ijk}\left( n_{nb}^{\pm }\right)
_{j}\left( n_{na}^{\pm }\right) _{k})\\
\frac{\partial q_{n}^{ab}}{\partial \theta _{i}^{n\pm }}|_{\theta _{i}^{n\pm
}=0}&=&\frac{1}{2}i\gamma ^{2}\gamma ^{\pm }j_{na}j_{nb}\left( \left(
n_{na}^{\pm }\right) _{i}+\left( n_{nb}^{\pm }\right) _{i}\right) (1-\left(
n_{na}\right) _{j}\left( n_{nb}\right) _{j})
        \end{eqnarray}
         The second derivatives of the insertions
            \begin{eqnarray}
              \frac{\partial ^{2}q_{n}^{ab}}{\partial \theta _{j}^{a\pm }\partial \theta
_{i}^{a\pm }}|_{\theta =0}&=&\frac{1}{4}\gamma ^{2}\gamma ^{\pm }j_{na}j_{nb}(\left( n_{nb}^{\pm
}\right) _{j}\left( n_{na}^{\pm }\right) _{i}+\left( n_{nb}^{\pm }\right)
_{i}\left( n_{na}^{\pm }\right) _{j}-2\left( n_{nb}\right) _{r}\left(
n_{na}\right) _{r}\left( n_{na}^{\pm }\right) _{i}\left( n_{na}^{\pm
}\right) _{j} \\ \nonumber
&&-i\left( n_{nb}^{\pm }\right) _{k}\left( n_{na}^{\pm }\right) _{m}\left(
\varepsilon _{kmj}\left( n_{na}^{\pm }\right) _{i}+\varepsilon _{kmi}\left(
n_{na}^{\pm }\right) _{j}\right) )
        \end{eqnarray}
            \begin{eqnarray}
\frac{\partial ^{2}q_{n}^{ab}}{\partial \theta _{j}^{n\pm }\partial \theta
_{i}^{n\pm }}|_{\theta =0}&=&\frac{1}{4}\gamma ^{2}\gamma ^{\pm }j_{na}j_{nb}(2\left( n_{na}^{\pm
}\right) _{i}\left( n_{nb}^{\pm }\right) _{j}+2\left( n_{nb}^{\pm }\right)
_{i}\left( n_{na}^{\pm }\right) _{j} \\ \nonumber
&&-2\left( n_{nb}\right) _{r}\left( n_{na}\right) _{r}\left( n_{na}^{\pm
}\right) _{i}\left( n_{na}^{\pm }\right) _{j}-2\left( n_{na}\right)
_{r}\left( n_{nb}\right) _{r}\left( n_{nb}^{\pm }\right) _{i}\left(
n_{nb}^{\pm }\right) _{j} \\ \nonumber
&&-i\left( n_{nb}^{\pm }\right) _{k}\left( n_{na}^{\pm }\right) _{m}\left(
\varepsilon _{kmj}\left( n_{na}^{\pm }\right) _{i}+\varepsilon _{kmi}\left(
n_{na}^{\pm }\right) _{j}\right) \\ \nonumber
&&-i\left( n_{na}^{\pm }\right) _{k}\left( n_{nb}^{\pm }\right) _{m}\left(
\varepsilon _{kmj}\left( n_{nb}^{\pm }\right) _{i}+\varepsilon _{kmi}\left(
n_{nb}^{\pm }\right) _{j}\right) ) \\ \nonumber
&&-\frac{1}{4}\gamma ^{2}\left( \gamma ^{\pm }\right)
^{2}j_{na}j_{nb}((\delta _{li}-\left( n_{na}^{\pm }\right) _{l}\left(
n_{na}^{\pm }\right) _{i}-i\varepsilon _{lik}\left( n_{na}^{\pm }\right)
_{k}) \\ \nonumber
&&\times (\delta _{lj}-\left( n_{nb}^{\pm }\right) _{l}\left( n_{nb}^{\pm
}\right) _{j}-i\varepsilon _{ljm}\left( n_{nb}^{\pm }\right) _{m}) \\ \nonumber
&&+(\delta _{li}-\left( n_{nb}^{\pm }\right) _{l}\left( n_{nb}^{\pm }\right)
_{i}-i\varepsilon _{lik}\left( n_{nb}^{\pm }\right) _{k})(\delta
_{lj}-\left( n_{na}^{\pm }\right) _{l}\left( n_{na}^{\pm }\right)
_{j}-i\varepsilon _{ljm}\left( n_{na}^{\pm }\right) _{m}))
     \end{eqnarray}
            \begin{eqnarray}
\frac{\partial ^{2}q_{n}^{ab}}{\partial \theta _{j}^{n\pm }\partial \theta
_{i}^{n\mp }}|_{\theta =0}&=&-\frac{1}{4}\gamma ^{2}\gamma ^{\pm }\gamma ^{\mp }j_{na}j_{nb}((\delta
_{li}-\left( n_{na}^{\mp }\right) _{l}\left( n_{na}^{\mp }\right)
_{i}-i\varepsilon _{lik}\left( n_{na}^{\mp }\right) _{k}) \\ \nonumber
&&\times (\delta _{lj}-\left( n_{nb}^{\pm }\right) _{l}\left( n_{nb}^{\pm
}\right) _{j}-i\varepsilon _{ljm}\left( n_{nb}^{\pm }\right) _{m}) \\ \nonumber
&&+(\delta _{li}-\left( n_{nb}^{\mp }\right) _{l}\left( n_{nb}^{\mp }\right)
_{i}-i\varepsilon _{lik}\left( n_{nb}^{\mp }\right) _{k})(\delta
_{lj}-\left( n_{na}^{\pm }\right) _{l}\left( n_{na}^{\pm }\right)
_{j}-i\varepsilon _{ljm}\left( n_{na}^{\pm }\right) _{m}))
     \end{eqnarray}
            \begin{eqnarray}
\frac{\partial ^{2}q_{n}^{ab}}{\partial \theta _{j}^{n\pm }\partial \theta
_{i}^{a\pm }}|_{\theta =0}
&=&\frac{1}{2}\gamma ^{2}\gamma ^{\pm }j_{na}j_{nb}(\left( n_{nb}\right)
_{r}\left( n_{na}\right) _{r}\left( n_{na}^{\pm }\right) _{i}\left(
n_{na}^{\pm }\right) _{j}-\left( n_{nb}^{\pm }\right) _{i}\left( n_{na}^{\pm
}\right) _{j} \\ \nonumber
&&+i\left( \varepsilon _{ijk}\left( n_{nb}^{\pm }\right) _{k}+\left(
n_{na}^{\pm }\right) _{i}\varepsilon _{jkl}\left( n_{na}^{\pm }\right)
_{k}\left( n_{nb}^{\pm }\right) _{l}-\left( n_{na}\right) _{r}\left(
n_{nb}\right) _{r}\varepsilon _{ijn}\left( n_{na}^{\pm }\right) _{n}\right) )
\\ \nonumber
&&+\frac{1}{4}\gamma ^{2}\left( \gamma ^{\pm }\right)
^{2}j_{na}j_{nb}(\delta _{li}-\left( n_{na}^{\pm }\right) _{l}\left(
n_{na}^{\pm }\right) _{i}-i\varepsilon _{lik}\left( n_{na}^{\pm }\right)
_{k}) \\ \nonumber
&&\times (\delta _{jl}-\left( n_{nb}^{\pm }\right) _{l}\left( n_{nb}^{\pm
}\right) _{j}-i\varepsilon _{ljm}\left( n_{nb}^{\pm }\right) _{m})
     \end{eqnarray}
            \begin{eqnarray}
\frac{\partial ^{2}q_{n}^{ab}}{\partial \theta _{j}^{n\mp }\partial \theta
_{i}^{a\pm }}|_{\theta =0}&=&\frac{1}{4}\gamma ^{2}\gamma ^{\pm }\gamma ^{\mp }j_{na}j_{nb}(\delta
_{li}-\left( n_{na}^{\pm }\right) _{l}\left( n_{na}^{\pm }\right)
_{i}-i\varepsilon _{lik}\left( n_{na}^{\pm }\right) _{k}) \\ \nonumber
&&\times (\delta _{lj}-\left( n_{nb}^{\mp }\right) _{l}\left( n_{nb}^{\mp
}\right) _{j}-i\varepsilon _{ljk}\left( n_{nb}^{\mp }\right) _{k})
     \end{eqnarray}
            \begin{eqnarray}
\frac{\partial ^{2}q_{n}^{ab}}{\partial \theta _{j}^{b\pm }\partial \theta
_{i}^{a\pm }}|_{\theta =0}&=&-\frac{1}{4}\gamma ^{2}\left( \gamma ^{\pm }\right)
^{2}j_{na}j_{nb}(\delta _{li}-\left( n_{na}^{\pm }\right) _{l}\left(
n_{na}^{\pm }\right) _{i}-i\varepsilon _{lik}\left( n_{na}^{\pm }\right)
_{k}) \\ \nonumber
&&\times (\delta _{lj}-\left( n_{nb}^{\pm }\right) _{l}\left( n_{nb}^{\pm
}\right) _{j}-i\varepsilon _{ljk}\left( n_{nb}^{\pm }\right) _{k})
     \end{eqnarray}
            \begin{eqnarray}
\frac{\partial ^{2}q_{n}^{ab}}{\partial \theta _{j}^{b\mp }\partial \theta
_{i}^{a\pm }}|_{\theta =0}
&=&-\frac{1}{4}\gamma ^{2}\gamma ^{\pm }\gamma ^{\mp }j_{na}j_{nb}(\delta
_{li}-\left( n_{na}^{\pm }\right) _{l}\left( n_{na}^{\pm }\right)
_{i}-i\varepsilon _{lik}\left( n_{na}^{\pm }\right) _{k}) \\
 \nonumber
&&\times (\delta _{lj}-\left( n_{nb}^{\mp }\right) _{l}\left( n_{nb}^{\mp
}\right) _{j}-i\varepsilon _{ljk}\left( n_{nb}^{\mp }\right) _{k})
            \end{eqnarray}

\section{Numerical comparison}\label{numerical}
Let us write some explicit terms of the loop three-point  function.
  \begin{equation}
  \begin{split}
    G_{123}^{444444}&=-\frac{3}{20} \sqrt{\frac{3}{5}} A^4 (151 \beta_0^3+\beta_0^2
   (504 \beta_2-817 \beta_1)+\beta_0 \left(1473 \beta_1^2-1832
   \beta_1 \beta_2+532 \beta_2^2\right)\\
   &\quad-959 \beta_1^3+1944
   \beta_1^2 \beta_2-1308 \beta_1 \beta_2^2+312
   \beta_2^3)
  \end{split}
  \end{equation}
  \begin{equation}
    \begin{split}
      G_{123}^{444445}&=-\frac{1}{20 \sqrt{15}}A^4 (3667 \beta_0^3+\beta_0^2 (10578 \beta_2-20099
  \beta_1)+\beta_0 \left(37161\beta_1^2-39964\beta_1
   \beta_2+10844 \beta_2^2\right)\\
   &\quad-22473\beta_1^3+36138
  \beta_1^2 \beta_2-19116\beta_1 \beta_2^2+3264
   \beta_2^3)
    \end{split}
  \end{equation}
  \begin{equation}
    \begin{split}
      G_{123}^{444455}&=-\frac{1}{10} \sqrt{\frac{3}{5}} A^4 (24 \beta_0^3+\beta_0^2
   (41 \beta_2-203\beta_1)+2 \beta_0 \left(71\beta_1^2+171
  \beta_1 \beta_2-116 \beta_2^2\right)\\
   &\quad+69\beta_1^3-639
  \beta_1^2 \beta_2+648\beta_1 \beta_2^2-192
   \beta_2^3)
    \end{split}
  \end{equation}
\end{widetext}
Here the parameters $\beta _{0},\beta _{1},\beta _{2}$ are directly related to the parameters  $\alpha_k$ appearing in the boundary state (see Eqs.(197-109) in \cite{Bianchi:2009ri}).  If we use for these, as an example, the values of these parameters given in \cite{Bianchi:2009ri}:
\begin{equation}\label{C}
\beta _{0}=0,\ \ \ \ \ \beta _{1}=-\frac{1}{2304\sqrt{3}},\ \ \ \ \ \beta
_{2}=-\frac{7}{9216\sqrt{3}}.
\end{equation}
Then we obtain
  \begin{eqnarray*}
    G_{123}^{444444}&=&-\frac{35 \sqrt{5} A^4}{391378894848}\\
    G_{123}^{444445}&=&\frac{5 \sqrt{5} A^4}{195689447424}\\
    G_{123}^{444455}&=&\frac{5 \sqrt{5} A^4}{97844723712}
  \end{eqnarray*}
On the other hand, from perturbation theory, we  get
  \begin{eqnarray*}
    (G_{123}^{444444})_{QFT}&=&-\frac{3}{16384000}C,\\
    (G_{123}^{444445})_{QFT}&=&\frac{13}{65536000}C,\\
    (G_{123}^{444455})_{QFT}&=&\frac{11}{16384000}C.
  \end{eqnarray*}
  where $C$ is a constant.
  The ratios of the two give
  \begin{eqnarray*}
    \frac{G_{123}^{444444}}{(G_{123}^{444444})_{QFT}}/\frac{G_{123}^{444445}}{(G_{123}^{444445})_{QFT}}&=&\frac{91}{24},\\
    \frac{G_{123}^{444444}}{(G_{123}^{444444})_{QFT}}/\frac{G_{123}^{444455}}{(G_{123}^{444455})_{QFT}}&=&\frac{77}{12},\\
  \end{eqnarray*}
which do not match. The use of the values for the $\beta$ coefficients given in Eq.\eqref{C} is of course rather questionable, and should not be taken too seriously. The main purpose of this computation is to show that the expectation values can indeed be computed completely explicitly.

\vfill

\bibliographystyle{apsrev4-1}
  \bibliography{BiblioCarlo,3P}
\end{document}